\documentclass[
12pt,
superscriptaddress,
 amsmath,amssymb,
aps,
]{revtex4-2}

\pdfoutput=1

\usepackage{graphicx}

\usepackage{subfigure}

\usepackage{epstopdf}

\usepackage{xcolor}

\usepackage{commath}

\usepackage{amsmath}

\usepackage{gensymb}

\usepackage{upgreek}

\usepackage{dcolumn}
\usepackage{bm}

\begin{document}

\preprint{APS/123-QED}

\title{Monopole density and antiferromagnetic domain control in spin-ice iridates}

\author{M. J. Pearce}
\affiliation{Department of Physics, University of Warwick, Coventry, CV4 7AL, UK.}
\author{K. G{\"o}tze}
\affiliation{Department of Physics, University of Warwick, Coventry, CV4 7AL, UK.}
\author{A. Szab{\'o}}
\affiliation{T.C.M. Group, Cavendish Laboratory, J. J. Thomson Avenue, University of Cambridge, Cambridge, CB3 0HE, UK.}
\author{T. S. Sikkenk}
\affiliation{T.C.M. Group, Cavendish Laboratory, J. J. Thomson Avenue, University of Cambridge, Cambridge, CB3 0HE, UK.}
\affiliation{Institute for Theoretical Physics and Center for Extreme Matter and Emergent Phenomena, Utrecht University, Leuvenlaan 4, 3584 CE Utrecht, The Netherlands.}
\author{M. R. Lees}
\affiliation{Department of Physics, University of Warwick, Coventry, CV4 7AL, UK.}
\author{A.~T.~Boothroyd}
\affiliation{Department of Physics, University of Oxford, Clarendon Laboratory, Oxford, OX1 3PU, UK.}
\author{D. Prabhakaran}
\affiliation{Department of Physics, University of Oxford, Clarendon Laboratory, Oxford, OX1 3PU, UK.}
\author{C. Castelnovo}
\email{cc726@cam.ac.uk}
\affiliation{T.C.M. Group, Cavendish Laboratory, J. J. Thomson Avenue, University of Cambridge, Cambridge, CB3 0HE, UK.}
\author{P. A. Goddard}
\email{p.goddard@warwick.ac.uk}
\affiliation{Department of Physics, University of Warwick, Coventry, CV4 7AL, UK.}

\date{\today}

%%%%%%%%%%%%%%%%%%%%%%%%%%%%%%%%%%%%%%%%%%%%%%%%%%%%%%%%%%%%%%%%%%%%%%%%%%%%%%%%%%%%%%%%%%%%%%%%%%

\begin{abstract}
Frustration in magnetic systems~\cite{LMM11book} is fertile ground for complex behaviour, including unconventional ground states with emergent symmetries, topological properties, and exotic excitations~\cite{Moessner06, Balents10, Knolle19}. A canonical example is the emergence of magnetic-charge-carrying quasiparticles in spin-ice compounds~\cite{Bramwell01, Castelnovo12}. Despite extensive work, a reliable experimental indicator of the density of these magnetic monopoles in spin-ice systems is yet to be found. Here, using measurements on single crystals of Ho$_{2}$Ir$_{2}$O$_{7}$ in combination with dipolar Monte Carlo simulations, we show that the magnetoresistance is highly sensitive to the density of monopoles. Moreover, we find that for the orientations of magnetic field in which the monopole density is enhanced, a strong coupling emerges between the magnetic charges on the holmium sublattice and the antiferromagnetically ordered iridium ions, leading to an ability to manipulate the antiferromagnetic domains via a uniform external field. Our results pave the way to a quantitative experimental measure of monopole density and provide a powerful illustration of the interplay between the various magnetic and electronic degrees of freedom in the frustrated pyrochlore iridates. This interdependence holds promise for potential functional properties arising from the link between magnetic and electric charges~\cite{Zhang20}, as well as for the control of antiferromagnetic domain walls, a key goal in the design of next-generation spintronic devices~\cite{Baltz18}.
\end{abstract}

%%%%%%%%%%%%%%%%%%%%%%%%%%%%%%%%%%%%%%%%%%%%%%%%%%%%%%%%%%%%%%%%%%%%%%%%%%%%%%%%%%%%%%%%%%%%%%%%%%

\maketitle

\section{\label{sec:level1}Introduction}

 Quintessential examples of spin-ice compounds include Ho$_{2}$Ti$_{2}$O$_{7}$ and Dy$_{2}$Ti$_{2}$O$_{7}$~\cite{Harris97,Bramwell01,Castelnovo12}, where the magnetic Ho$^{3+}$/Dy$^{3+}$ ions sit at the vertices of corner-sharing tetrahedra, which connect to form a pyrochlore lattice. The rare-earth moments are constrained by the easy-axis anisotropy to point either into, or out of, each tetrahedron. The resultant ground state is one where two spins point into each tetrahedron and two point out (2I2O), referred to as the `ice rule' by analogy with the proton disorder in water ice~\cite{Pauling35}. The natural excitations of a spin-ice system arise from flipping a moment, resulting in one tetrahedron exhibiting a one-in-three-out spin configuration (1I3O), and the neighbouring tetrahedron arranged three-in-one-out (3I1O)~\cite{Fulde02, Moessner03, Ryzhkin05}. These ice-rule violating defects can be separated by flipping a chain of spins, and are understood as deconfined positive and negative magnetic monopoles, respectively~\cite{Castelnovo08, Morris09, Fennell09, Kadowaki09}.

\begin{figure}[t]
\centering
\vspace{-12mm}
\hspace{-11mm}
\includegraphics[width=0.54\textwidth]{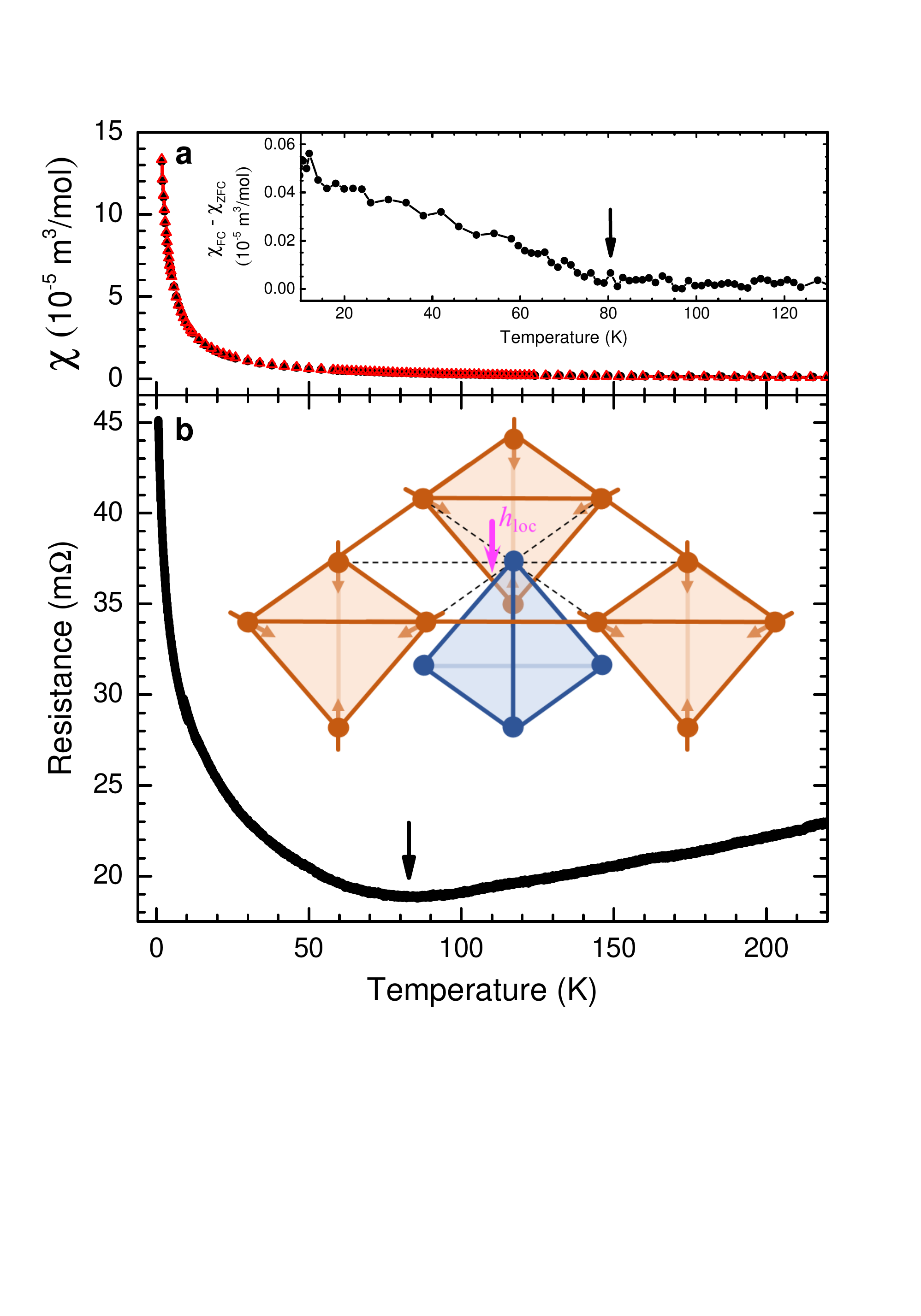}
\vspace{-30mm}
\caption{\textbf{Iridium ordering and metal-insulator transition in single-crystalline Ho$_{2}$Ir$_{2}$O$_{7}$}. (a) Field-cooled (red open triangles) and zero-field-cooled (black filled circles) magnetic susceptibility measured in a 0.01~T [100] magnetic field. Inset: Field-cooled magnetic susceptibility with the zero-field-cooled measurement subtracted. The ordering of the Ir moments at approximately 80~K is indicated by the arrow. (b) Resistance measured under zero applied field, exhibiting a metal-insulator transition at approximately 80~K as indicated by the arrow. Inset: Schematic of the pyrochlore structure of Ho$_{2}$Ir$_{2}$O$_{7}$, in which each Ho$^{3+}$ ion (blue) has six Ir$^{4+}$ (orange) nearest neighbours. The long range magnetic order of the Ir moments (orange arrows) results in a net local effective field at the Ho sites (${\bf h}_{\rm loc}$) aligned either parallel or antiparallel to the local [111] direction (indicated here by the magenta arrow for the uppermost Ho site)~\cite{Lefrancois17}.}
\label{MIT}
\end{figure}

In contrast with the non-magnetic Ti atoms of RE$_{2}$Ti$_{2}$O$_{7}$ (RE = rare earth), pyrochlore iridates (RE$_{2}$Ir$_{2}$O$_{7}$), contain the additional magnetism of the Ir$^{4+}$ ions, which sit on a pyrochlore lattice interpenetrating that of the rare-earth moments. The Ir moments are coupled by antiferromagnetic exchange interactions and, with the exception of Pr$_{2}$Ir$_{2}$O$_{7}$, spontaneously order at sufficiently low temperatures such that they alternate pointing all into/all out of adjacent tetrahedra on the pyrochlore lattice~\cite{Donnerer16}. This is observed as a slight bifurcation of the field-cooled and zero-field cooled magnetic susceptibilities~\cite{Matsuhira11}, which for Ho$_{2}$Ir$_{2}$O$_{7}$ are otherwise dominated by the paramagnetism of the large Ho moments (Figure~1a). Pyrochlore iridates also undergo a metal-insulator transition which occurs concomitantly with the Ir ordering~\cite{Matsuhira11}. Measurements of the magnetic susceptibility and resistance show that for single crystals of Ho$_{2}$Ir$_{2}$O$_{7}$, both the Ir ordering (Figure~1a) and the metal-insulator transition (Figure~1b) occur at approximately 80~K. The ordered Ir moments produce a local effective magnetic field (${\bf h}_{\rm loc}$) at the Ho sites aligned either parallel or antiparallel to the local $\langle$111$\rangle$ directions (see inset to Figure~\ref{MIT}b). These fields, when combined with the spin-ice physics, lead to a finite concentration of monopoles on the Ho sublattice~\cite{Lefrancois17}.

Spin ices exhibit an anisotropic response to externally applied magnetic fields. A [100] field stabilises a specific monopole-free 2I2O ground state, whereas a sufficiently large [111] field promotes the formation of magnetic monopoles, arranging the rare-earth moments into a 3I1O/1I3O monopole crystal. The experimental study of this anisotropic behaviour requires measurements on single crystals, something which has only become possible very recently for the pyrochlore iridates~\cite{Cathelin20}.

Here we present measurements of the magnetic field dependence of the magnetisation and resistance of Ho$_{2}$Ir$_{2}$O$_{7}$ alongside corresponding dipolar Monte Carlo simulations of the magnetisation and monopole density. To the best of our knowledge these are the first such experimental measurements on single crystals of this material. We find that the form of the magnetoresistance is strongly linked to the density of magnetic monopoles, suggesting a route to quantitatively measure the latter. Moreover, our results show that applying a [111] magnetic field manipulates the ratio of the antiferromagnetic Ir domains in a highly robust, reproducible and controlled manner, potentially offering a new avenue for antiferromagetic spintronic applications.

%%%%%%%%%%%%%%%%%%%%%%%%%%%%%%%%%%%%%%%%%%%%%%%%%%%%%%%%%%%%%%%%%%%%%%%%%%%%%%%%%%%%%%%%%%%%%%%%%%

\section{\label{sec:level1}Results and Discussion}

%%%%%%%%%%%%%%%%%%%%%%%%%%%%%%%%%%%%%%%%%%%%%%%%%%%%%%%%%%%%%%%%%%%%%%%%%%%%%%%%%%%%%%%%%%%%%%%%%%

\subsection{\label{sec:level2}Applied magnetic field parallel to [100]}

\begin{figure*}[]
\includegraphics[width=0.50\textwidth]{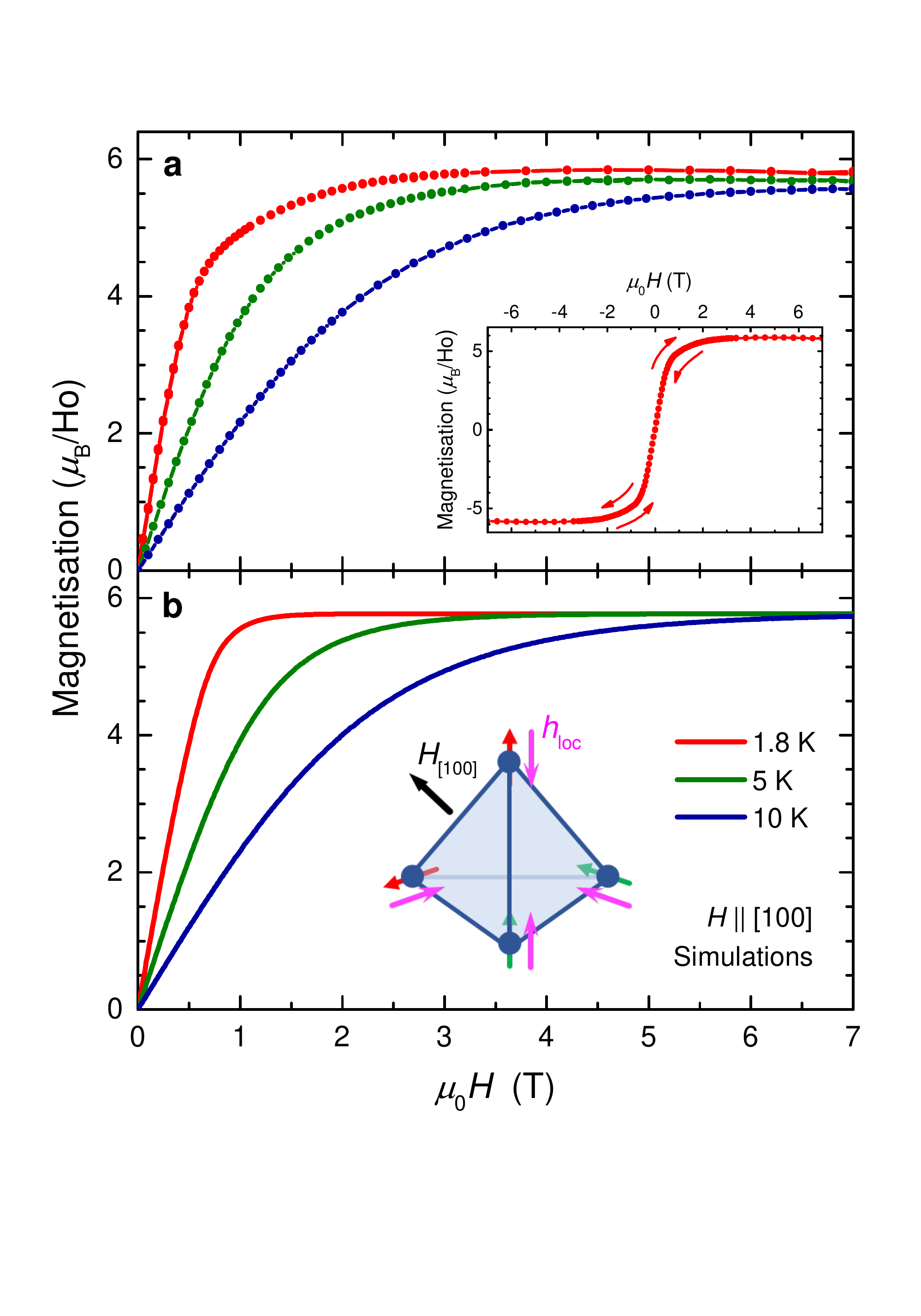}
\hspace{-7mm}
\includegraphics[width=0.50\textwidth]{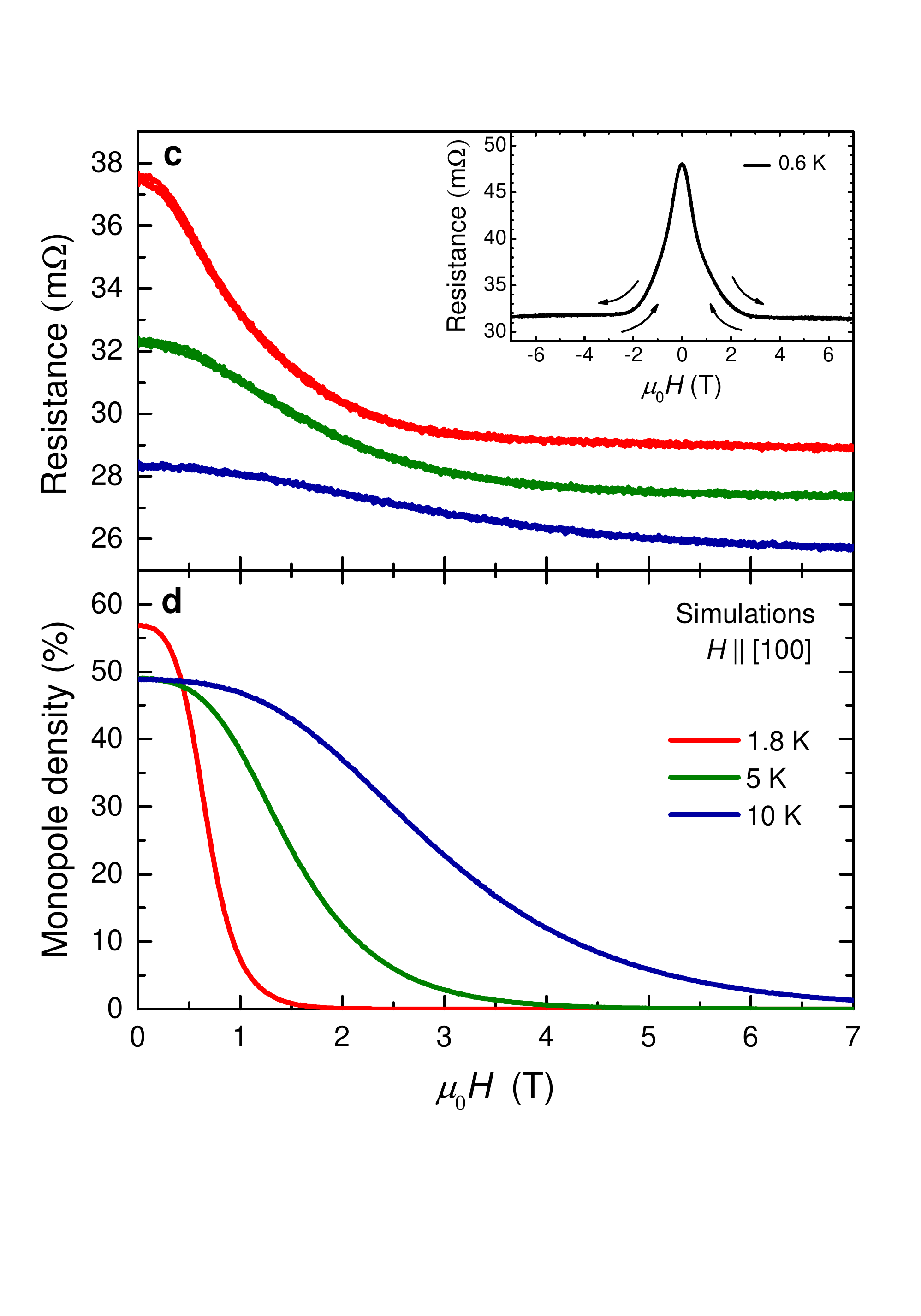}
\vspace{-20mm}
\caption{\textbf{Ho$_{2}$Ir$_{2}$O$_{7}$ under the application of a [100] magnetic field.} (a) Measurements and (b) Monte Carlo simulations of the magnetisation at various temperatures. The inset to (a) shows a measurement of the magnetisation at 1.8~K upon sweeping the field continuously between the positive and negative field limits. The inset to (b) shows a single tetrahedron of the Ho$^{3+}$ sublattice. Magenta arrows indicate the local effective field ${\bf h}_{\rm loc}$ due to the ordered Ir moments. Under an externally applied [100] magnetic field (black arrow) the Ho moments order into a 2I2O configuration, oriented either parallel (green arrow) or antiparallel (red arrow) to ${\bf h}_{\rm loc}$. (c) Resistance measurements and (d) Monte Carlo simulations of the density of single monopoles at various temperatures. The inset to (c) shows a measurement of the resistance at 0.6~K upon sweeping the field continuously between the positive and negative field limits. The negative magnetoresistance results from a combination of the paramagnetic response of spin ice at fixed monopole density and variations in the monopole density via the mechanisms described in the text.}
\label{001}
\end{figure*}

First we report the results of magnetisation and resistance measurements on single crystals of Ho$_{2}$Ir$_{2}$O$_{7}$ under an externally applied magnetic field oriented along the [100] crystallographic direction. The phenomenology for this orientation allows a better understanding and appreciation of the rich and complex physics at play when a field is applied along the [111] direction, presented in the following section.

Figure~\ref{001}a shows the magnetisation rising rapidly under an applied [100] field to a saturation magnetisation, $M_{[100]}^{\rm sat}$, of (5.8$~\pm$~0.4)~$\mu_{\rm B}$/Ho. This is in excellent agreement with the value expected from the ice rules for this orientation of 5.77~$\mu_{\rm B}$/Ho~\cite{Harris98, Petrenko03}. We note that on sweeping the field continuously between the positive and negative field limits no hysteresis between the upsweeps and downsweeps is observed (see inset to Figure~\ref{001}a). Monte Carlo simulations of the magnetisation were performed as described in the methods section. Figure~\ref{001}b shows the simulations for $H\parallel$[100] which quantitatively reproduce the behaviour of the measured magnetisation curves over the studied temperature range.

Figure~\ref{001}c presents the evolution of the electrical resistance under an applied [100] magnetic field. From a temperature-dependent starting value, there is an initial large negative magnetoresistance which flattens out at higher fields. As with the magnetisation curves, a continuous sweep of the field between the positive and negative field limits yields negligible hysteresis (see inset to Figure~\ref{001}c). 

The temperature dependence of the zero-field resistance arises due to the insulating nature of Ho$_{2}$Ir$_{2}$O$_{7}$ below the metal-insulator transition at 80~K. The negative magnetoresistance is caused by a reduction in scattering as the Ho moments order under an applied magnetic field. This originates in part from the paramagnetic response of spin ice at fixed monopole density and in part from the suppression of the density of magnetic monopoles at sufficiently large field values. Indeed, Figure~\ref{001}d presents Monte Carlo simulations which show how a [100] field acts to suppress the monopole density as the Ho moments order into a 2I2O monopole-free magnetic state. The mechanisms linking resistance and monopole density will be discussed further in the next section (see also Supplementary Section S1).

As the temperature is increased to 10~K (Figure~\ref{001}c), the initial negative magnetoresistance broadens out to higher fields. This is consistent with the temperature dependence of the above two contributions. As the temperature is increased a larger field is required to order the Ho moments (Figure~\ref{001}a) and thus the reduction in spin scattering occurs over a broader field range. Likewise, higher fields are required to achieve the monopole-free 2I2O state associated with saturation in this orientation, resulting in a slower reduction of the monopole density (Figure~\ref{001}d). This behaviour continues for measurements of the magnetisation and resistance at temperatures higher than 10~K (see Supplementary Section S2).

%%%%%%%%%%%%%%%%%%%%%%%%%%%%%%%%%%%%%%%%%%%%%%%%%%%%%%%%%%%%%%%%%%%%%%%%%%%%%%%%%%%%%%%

\subsection{\label{Sec: 111}Applied magnetic field parallel to [111]}

\begin{figure}
\vspace{-5mm}
\includegraphics[width=\textwidth]{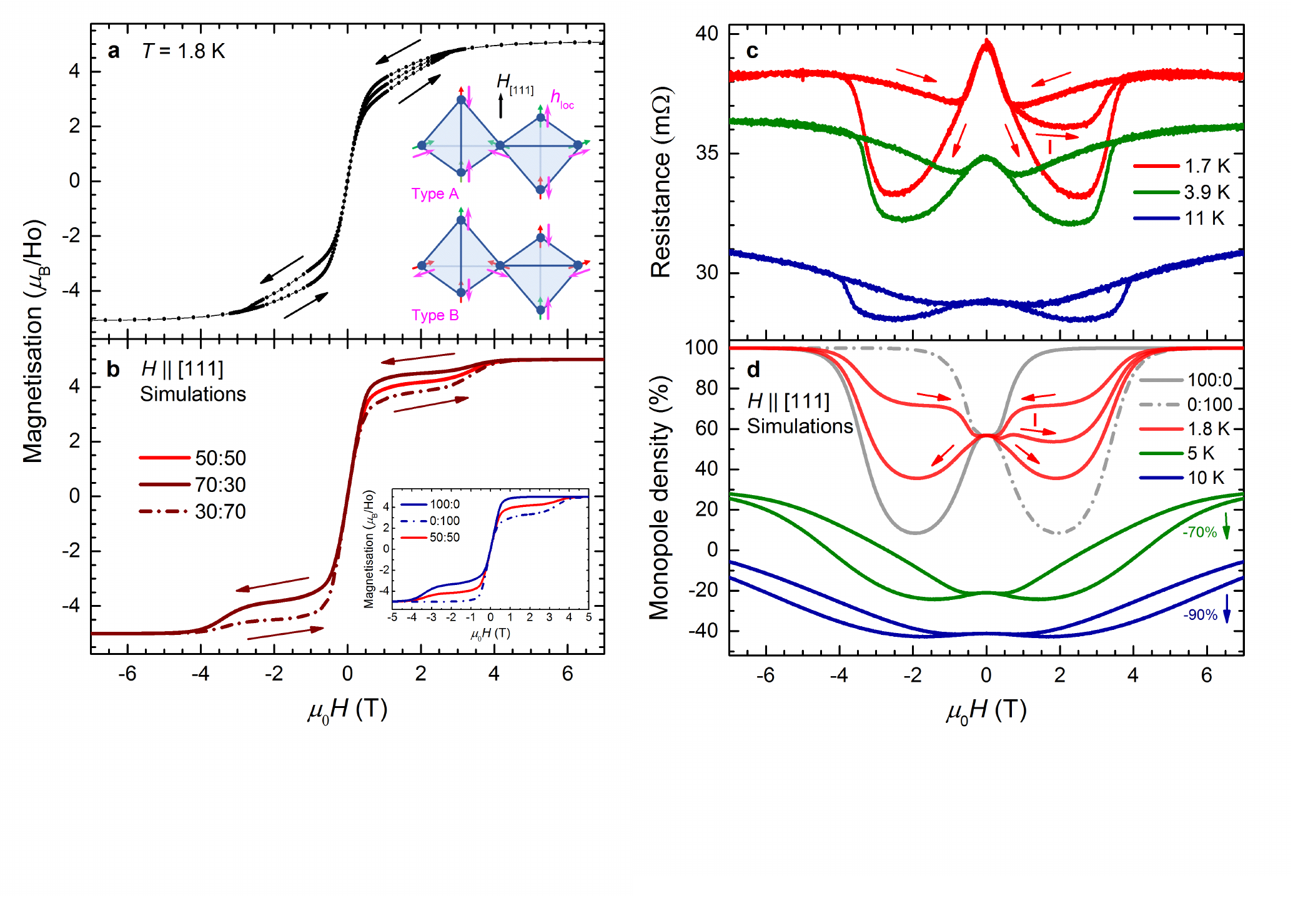}
\vspace{-35mm}
\caption{\textbf{Ho$_{2}$Ir$_{2}$O$_{7}$ under the application of a [111] magnetic field.} (a) Measurements and (b) Monte Carlo simulations of the magnetisation at 1.8~K. The direction of the hysteresis observed experimentally is given by the arrows in (a); the additional curve lying between the upsweep and downsweep for $H >$ 0 is the virgin curve (initial sweep after cooling the sample in zero magnetic field). The inset to (a) depicts tetrahedra of the Ho$^{3+}$ sublattice. Magenta arrows indicate the local effective field ${\bf h}_{\rm loc}$ due to the ordered Ir moments for a type-A (top) and type-B (bottom) domain. Under an external [111] magnetic field (black arrow) the Ho moments realise a 3I1O/1I3O monopole crystal, orienting either parallel (green arrows) or antiparallel (red arrows) to ${\bf h}_{\rm loc}$. For this direction of applied field, the alignment favours type-A Ir domains.
The 100:0 and 0:100 curves in the inset to (b) are simulations for a type-A and type-B single-domain crystal, respectively. Other curves in (b) are weighted averages of these single-domain curves. Plastic deformation of the Ir domain ratio occurs due to the Ho--Ir interaction; the magnetisation of the Ho moments follows the 50:50 curve for the virgin sweep and then the 70:30/30:70 curves as indicated by the arrows. (c) Resistance measurements and (d) Monte Carlo simulations of the density of single monopoles. The 100:0/0:100 curves are simulations for a type~A/B single-domain crystal. The monopole density is simulated using a 30:70/70:30 average of the two Ir domain types for increasing/decreasing fields and a 50:50 average for the virgin curve (1.8~K only, labelled `\textsf{I}') as indicated by the arrows. The 5~K/10~K curves in (d) have been shifted down by 70\%/90\% for clarity. Higher temperature magnetisation and resistance data are presented in Supplementary Section S3.}
\label{111}
\end{figure}

Figure~\ref{111}a shows that under the application of a [111] magnetic field, the magnetisation rises to saturation at $M_{[111]}^{\rm sat}~=~(5.1~\pm~0.4)~\mu_{\rm B}$/Ho, in excellent agreement with the expected value of 5.0~$\mu_{\rm B}$/Ho for a 3I1O/1I3O spin configuration~\cite{Harris98, Petrenko03}.

A striking observation is the presence of hysteresis in the magnetisation for this orientation. Notably, the hysteresis is closed at zero applied field, indicating that the field sweeps are slow enough for the Ho moments to remain in equilibrium. The hysteresis opens at a finite value of the applied field: for example, at 1.8~K the hysteresis becomes appreciable around 0.2~T and then closes at 2.9~T~\cite{Note2}. The virgin curve, defined as the first field sweep after cooling from above the metal-insulator transition in zero magnetic field, sits between the subsequent downsweeps and upsweeps of the hysteresis loop. Other than the virgin sweep, the positive and negative quadrants of the magnetisation loop are symmetric within experimental error. 

In the Monte Carlo simulations we attempted to induce hysteretic behaviour by increasing the rate at which the magnetic field was ramped, but this necessarily resulted in a hysteresis that was open at $H = 0$, in contrast to the experimentally observed behaviour. Since single spin flip Monte Carlo dynamics appears to describe spin-ice materials accurately in the dipolar spin-ice model~\cite{Jaubert09}, this finding suggests that the origin of the hysteresis is not intrinsic to the Ho moments and their interactions: a conclusion supported by the fact that the experimental results are largely independent of sweep rate (see Supplementary Section S4). We are however able to explain the observed hysteresis by proposing that applying a [111] magnetic field plastically changes the ratio of Ir-domain types during the course of a field sweep, as we outline below. 

For the Ir order that exists below 80~K there are two possible domains, which we denote type~A and type~B~\cite{Ma15, Footnote1}. The diagrams in Figure~\ref{111}(a) show the arrangement at the Ho sites of the local effective field produced by the ordered Ir moments for both domain types. The two domains differ by a time-reversal transformation, corresponding to the reversal of the local effective field at each Ho site. An external magnetic field applied along the [111] crystallographic direction will push the initially disordered Ho moments into a 3I1O/1I3O monopole crystal at saturation~\cite{Harris98, Petrenko03}. This arrangement of Ho moments produces a relative energy saving for a type-A Ir domain due to the favourable alignment of the local fields, but leads to a relative energy cost for a type-B Ir domain. By contrast, the application of an external [$\bar{1}\bar{1}\bar{1}$] magnetic field will push the Ho moments in the opposite direction and hence favour type-B Ir domains.

If the crystal were to consist of a single domain then this effect would lead to an asymmetric magnetisation, as shown in the inset to Figure~\ref{111}b, where the 100:0 curve is the calculated magnetisation for a crystal of type-A Ir order only.  In this case the external [111] field works in tandem with the local Ir field and the Ho moments saturate rapidly into a 3I1O/1I3O monopole crystal. A [$\bar{1}\bar{1}\bar{1}$] field  ($H<0$ in the inset) applied to this domain competes with the local Ir field and so must be swept to a higher value to fully align the Ho ions. Also, because of this competition, the Ho moments rearrange via an intermediate regime with vanishing monopole density, resulting in a plateau in the magnetisation prior to saturation. The 0:100 curve shows the converse behaviour for a single-domain crystal of type-B Ir order. An average of these two lines is shown in the 50:50 curve, which simulates the response of a crystal containing a fixed and equal ratio of both Ir domains, and is symmetric but not hysteretic. It is clear that none of these situations alone can account for the observed symmetric and hysteretic behaviour; plastic deformation of domain boundaries must be considered. 

For the measured magnetisation in Figure~\ref{111}a the starting ratio of Ir domain types is expected to be approximately 50:50, because the sample is initially zero-field cooled through the ordering temperature. It is therefore reasonable that the virgin field sweep closely resembles the 50:50 simulation seen in Figure~\ref{111}b for $H>0$. As the applied [111] magnetic field is swept towards 7~T, an energetic pressure is exerted on the Ir domain walls. This acts to deform the Ir domains, skewing the ratio in favour of type A, reaching a maximum value once the Ho moments saturate (see Supplementary Section S5). As the applied field is swept from $+7$~T to 0~T the energetic pressure is relieved but the A:B ratio does not appear to change, suggesting a plastic deformation has taken place. On sweeping from 0~T to $-7$~T the influence on the Ir domain walls is in the opposite direction, skewing the ratio now in favour of type-B Ir domains. The ratio again remains fixed as the field is swept from $-7$~T to 0~T, before favouring type A once more as the field is increased along [111] and the hysteresis loop is completed. We note that our proposed mechanism is also consistent with the recent experimental results observed for Dy$_2$Ir$_2$O$_7$~\cite{Cathelin20}. 

In support of this picture, we find that subsequent to the virgin curve, the experimental data can be modelled well by a weighted average of the simulated magnetisation for type-A and type-B single-domain crystals in the ratio 70:30 and 30:70 for $+7$~T to $-7$~T and $-7$~T to $+7$~T, respectively (see Figure~\ref{111}b). This implies that the energetic pressure due to the saturated Ho moments does not force the Ir moments into a single ordered domain at high fields. The most likely explanation is that some form of disorder introduces a distribution of pinning energies for the Ir domain walls (see Supplementary Sections S6 and S7). This limits the imbalance of the Ir domain ratio reached, which the comparison between numerical simulations and experiments suggests is approximately 70:30. We note that in our approximate averaging procedure of the Monte Carlo simulations the Ir domain ratio changes abruptly at saturation, whereas experimentally we expect a gradual evolution during the field sweep (see Supplementary Section S7). This approximation likely accounts for the contrast between the smooth experimental curves and the step-like behaviour of the simulations.

As mentioned earlier, the application of a [100] magnetic field orders the Ho spins into a 2I2O spin-ice state. For this configuration, a tetrahedron in either Ir domain contains two Ho moments parallel to the local effective field and two that are antiparallel (see inset to Figure~\ref{001}b), giving rise to no energetic pressure between the two Ir orders. Consequently, no hysteresis is expected nor observed. A full study of the magnetisation and resistance under an applied [110] field is given in Supplementary Section S8, and provides additional validation for the mechanism proposed here. 

Figure~\ref{111}c shows the magnetoresistance in an applied [111] magnetic field which, like the magnetisation, is highly hysteretic. The virgin curve (\textsf{I}), shown only for the 1.7~K measurement, lies between the subsequent downsweeps and upsweeps. As with the [100] orientation, the form of the magnetoresistance is determined by: (i) the paramagnetic response of spin ice at fixed monopole density, which manifests as a marked drop in resistance at low fields; and (ii) variations of the monopole density as the field is swept. 

Figure~\ref{111}d presents calculations of the density of monopoles for $H\parallel[111]$. Under the application of a sufficiently large [111] magnetic field the Ho moments order into a 3I1O/1I3O monopole crystal and the monopole density rises to 100\%. Supported by the local effective field, this rise is rapid for the favourable Ir domain type (100:0 curve for $H > 0$), while the monopole density exhibits a minimum for the less favourable Ir domain (0:100 curve for $H > 0$), as the 3I1O/1I3O state is reached via a regime with vanishing monopole density. 

By analogy with the magnetisation analysis, weighted averages of the results for type-A and type-B single domain crystals are shown, where on moving around the hysteresis loop a 30:70 ratio represents sweeping the field from $-7$~T to $+7$~T, a 70:30 ratio $+7$~T to $-7$~T, and a 50:50 ratio the virgin curve (shown for 1.8~K only). Comparing measurements and simulations, it is clear that the resistance and monopole density share a consistent field and temperature dependence (a full discussion of the temperature dependence of both the magnetisation and resistance is given in Supplementary Section S3). The effect of the plastic deformation of the Ir domain walls is to introduce a notable hysteresis in the monopole density, which reproduces the hysteretic nature of the measured resistance extremely well. This indicates that the resistance of the material can be used as a reliable measure of the monopole density. 

We suggest that the resistance and monopole density are linked via two different scattering mechanisms that take place between the conduction electrons and the Ho magnetism at low temperatures. Magnetic scattering occurs between the electronic spin and the magnetic charge associated with a monopole~\cite{Castelnovo08}. Furthermore, lattice distortions due to the frustrated magnetic structure generate effective electric dipoles on each Ho tetrahedron hosting a monopole~\cite{Khomskii12}, which results in additional scattering of conduction electrons. Both mechanisms are charge-dipole type scattering from emergent monopoles, but via independent magnetic and electric channels. Both these effects lead to an electronic scattering rate, and hence a change in resistivity, which is proportional to the monopole density and sufficiently strong to account for the experimentally observed magnetoresistance (see Supplementary Section S1).

%%%%%%%%%%%%%%%%%%%%%%%%%%%%%%%%%%%%%%%%%%%%%%%%%%%%%%%%%%%%%%%%%%%%%%%%%%%%%%

\section{\label{sec:level1}Conclusions}

Since the early indirect evidence of magnetic monopoles in spin ice~\cite{Castelnovo08}, much effort has been devoted to their direct detection and characterisation, and in particular to measuring their density in experiments. The monopole density relates to both thermodynamic and dynamic properties of these materials, and several techniques have been proposed as a measurement proxy, for example specific heat~\cite{Morris09}, neutron scattering~\cite{Fennell09, Kadowaki09}, and magnetic susceptibility and noise measurements~\cite{Snyder04, Ryzhkin05, Jaubert09, Kaiser13, Kaiser15, Dusad19, Watson19}. Our results show that the form of the magnetoresistance of the spin ice Ho$_{2}$Ir$_{2}$O$_{7}$ is strongly linked to the concentration of magnetic monopoles, in a way that holds promise to develop a readily measurable and versatile experimental indicator of their density. 

Resistance measurements are a fast, straightforward, and widely available experimental technique, which can be performed on very small samples. They also permit time-resolved data collection over a wide temperature range and can be readily combined with high magnetic fields and applied pressure. We note that the scattering mechanisms linking the monopole density and resistance do not depend on the Ir magnetism. Consequently, this technique in principle lends itself to the study of magnetic monopoles in spin-ice systems irrespective of the presence or absence of magnetism at the transition metal site, provided the insulating band gap is sufficiently small to allow resistance measurements to take place. Physical pressure, chemical pressure, and strain in thin films offer routes to alter the band structure, potentially reducing the size of the insulating gap and thus widening the scope of compounds to which this technique may be applied. 

Our results demonstrate that the hysteresis in the magnetisation and resistance under an applied [111] magnetic field arises due to a plastic deformation of the antiferromagnetic Ir domain walls. Antiferromagnetic domains are a promising building block for future spintronic devices as they do not produce stray magnetic fields and possess ultrafast spin dynamics~\cite{Baltz18}. However, the manipulation of antiferromagnetic domain walls is challenging due to the net-zero magnetisation of the domains, and the staggered nature of the field required to interact with the alternating magnetic moments~\cite{Baltz18}. This is circumvented in Ho$_{2}$Ir$_{2}$O$_{7}$ by the interplay between the Ir domains and the frustrated ferromagnetism of the large Ho moments, which drives the motion of the domain walls. The result is a highly reproducible control over the antiferromagnetic domains via an external applied magnetic field. We also find that the domains are robust to low-level field noise and only fields in excess of approximately 1~T can perturb the magnetic microstructure (see Supplementary Section S5). Our results provide the key ingredients for new materials in which the control of antiferromagnetic domains is possible, namely (i) large frustrated moments; (ii) robust long-range antiferromagnetic order; and (iii) a strong coupling between (i) and (ii). 

We speculate that the plastic deformation of the Ir domains may be caused by a distribution of pinning energies for the domain walls (see Supplementary Section S7). While such plastic behaviour is generally expected for magnetic domain walls~\cite{Jiles15}, understanding and modelling it in these materials is an open and interesting question of direct relevance to potential applications, such as those in spintronics.

Finally we note that recent work on Bi$_{2}$Ir$_{2}$O$_{7}$/Dy$_{2}$Ti$_{2}$O$_{7}$ heterostructures has sought to address the grand challenge of converting spin excitations in a frustrated magnet into an electronic response by observing a connection between the spin states in the insulating titanate layer and the electronic properties of the non-frustrated iridate layer~\cite{Zhang20}. Our work brings to light the close links between the magnetic and electric charges in Ho$_{2}$Ir$_{2}$O$_{7}$, and hence establishes this interconnectedness in a single material. This, together with the inherent interplay between the Ir order and the frustrated ferromagnetism present in the spin-ice iridates, generates a rich and exciting playground for the study of complex and out-of-equilibrium behaviour and a new framework for possible future functional devices.

%%%%%%%%%%%%%%%%%%%%%%%%%%%%%%%%%%%%%%%%%%%%%%%%%%%%%%%%%%%%%%%%%%%%%%%%%%%%%

\section{Methods}

\subsection{Synthesis}

Phase-pure Ho$_{2}$Ir$_{2}$O$_{7}$ powder was prepared using high-purity ($>$99.99\%) Ho$_{2}$O$_{3}$ and IrO$_{2}$ with a molar ratio 1:1.05. The 5\% excess of IrO$_{2}$ was added to compensate for evaporation loss. The powder was thoroughly mixed with KF flux in the ratio 200:1 inside an Argon glove box and pressed into 15~mm diameter pellets. The pellets were placed inside a platinum crucible and sintered in a chamber furnace at 1100~\degree C for 10 hours, before cooling to 850~\degree C at 1~\degree C/hour and finally to room temperature at 60~\degree C/hour~\cite{Millican07}. Octahedral shaped single crystals were separated after dissolving the flux using hot water. Phase purity of the powder and single crystal samples was characterised using PANalytical and Supernova x-ray diffractometers, respectively. 

\subsection{Magnetometry Measurements}

The magnetisation (\textit{M}~vs~\textit{H}) and susceptibility (\textit{$\chi$}~vs~\textit{T}) of a single crystal of Ho$_{2}$Ir$_{2}$O$_{7}$ of mass $(0.134\pm 0.005)$~mg were measured using a Quantum Design MPMS superconducting quantum interference device (SQUID) magnetometer (with the exception of Supplementary Section S4). Magnetisation data at different field sweep rates (Supplementary Section S4) were collected using an Oxford Instruments vibrating sample magnetometer (VSM).

\subsection{Resistance Measurements}
Measurements of the electrical resistance of a single crystal of Ho$_{2}$Ir$_{2}$O$_{7}$ of approximate size 0.2~$\times$~0.2~$\times$~0.25~mm$^{3}$ were made using a four-wire technique with an 855~$\upmu$A ac current. Magnetic fields were applied using an Oxford Instruments superconducting magnet equipped with a $^{\rm 3}$He insert and were swept at a rate of 1~T/min.

\subsection{Simulations}

Monte Carlo (MC) simulations were performed using the full dipolar spin ice Hamiltonian~\cite{Hertog00} with an additional local [111] field to represent the coupling to the ordered iridium moments~\cite{Lefrancois17}:
\begin{multline}
    {\cal{H}} = \frac{J}3 \displaystyle \sum_{\langle ij\rangle} \sigma_i \sigma_j + D\ell^3 \sum_{ij}\sigma_i \sigma_j \left[\frac{\hat{\mathbf{e}}_i \cdot \hat{\mathbf{e}}_j }{r_{ij}^3} - \frac{3(\hat{\mathbf{e}}_i \cdot \mathbf{r}_{ij}) (\hat{\mathbf{e}}_j \cdot \mathbf{r}_{ij})}{r_{ij}^5}\right] 
    \\ \pm h_\mathrm{loc} \sum_i \sigma_i 
    - \mu_\mathrm{Ho} \mathbf{B} \cdot \sum_{i} \sigma_i \hat{\mathbf{e}}_i,
\end{multline}
where $J/k_\mathrm{B}=-1.56$~K~\cite{Bramwell01b} and $D/k_\mathrm{B}=1.34$~K~\cite{Lefrancois17} are the strengths of nearest-neighbour exchange and long-range dipolar interactions between holmium spins of magnitude $\mu_\mathrm{Ho}=10\,\mu_\mathrm{B}$~\cite{Lefrancois17}, respectively;
$\ell=3.6$~\AA{} is the distance between nearest-neighbour spins~\cite{Lefrancois17};
and $h_\mathrm{loc}/k_\mathrm{B}=3.5$~K is the net coupling to the ordered iridium moments, the sign of which depends on the domain type. 
The value of $h_\mathrm{loc}$ was chosen to ensure a good correspondence between the experimental and simulated magnetisation curves in the [100] direction: 
the difference between our value and the one used in Ref.~\cite{Lefrancois17} (6.3~K) likely derives from using the full dipolar Hamiltonian spin ice model instead of the nearest-neighbour approximation. For the range of fields and temperatures studied, we consider the Ir moments to be fixed in their all-in-all out local directions, similar to~\cite{Lefrancois17}.
The simulation was performed over $6\times6\times6$ cubic unit cells of the pyrochlore lattice (3\,456 spins); 
periodic boundary conditions for the dipolar interaction were enforced using Ewald summation,
including a demagnetising factor consistent with a spherical sample~\cite{Leeuw80}
to allow direct comparison with the (non-corrected) experimental data. 
A sweep rate of 0.2~Oe/MC~step was used for all simulations (which we expect to correspond to about 200~Oe/s in real time~\cite{Jaubert09});
we found that all simulations remained in thermodynamic equilibrium.

\bibliography{1mainPrePrint}

\begin{thebibliography}{39}
\bibitem [{Foo()}]{Footnote1}%
  \BibitemOpen
  \href@noop {} {}\bibinfo {note} {Elsewhere, including in {R}ef.~\cite{Ma15}
  which discusses {N}d$_{2}${I}r$_{2}${O}$_{7}$, the domain types are referred
  to as {AIAO} and {AOAI}}\BibitemShut {NoStop}%
\end{thebibliography}

%%%%%%%%%%%%%%%%%%%%%%%%%%%%%%%%%%%%%%%%%%%%%%%%%%%%%%%%%%%%%%%%%%%%%%%%%%%%%%%%%%%%%%%%%%%%%%%%%%

\section*{Acknowledgements}
\noindent We thank T. Orton for technical assistance. This project has received funding from the European Research Council (ERC) under the European Union's Horizon 2020 research and innovation program (Grant Agreement No. 681260). We acknowledge support from the Engineering and Physical Sciences Research Council (EPSRC) under the following grant numbers: EP/N509796/1 (M.J.P), EP/P034616/1 and EP/M007065/1 (C.C., T.S.S., and A.S.), and EP/N034872/1 and EP/J017124/1 (D.P. and A.T.B.).

\section*{Author Contributions}
\noindent D.P. and P.A.G. conceived the experiments. A.S. and C.C. conceived and performed the simulations, with initial involvement of T.S.S.. D.P. grew the samples. M.P., K.G., M.R.L., and P.A.G. performed the magnetisation and resistivity measurements and analysed the results.  M.P., P.A.G., A.S., and C.C. wrote the manuscript with input from all other co-authors. P.A.G., C.C., D.P. and A.T.B. supervised the project.

\end{document}

\preprint{APS/123-QED}

\title{Supplementary Information accompanying: 
\\ ``Monopole density and antiferromagnetic domain control in spin-ice iridates''}

\author{M. J. Pearce}
\affiliation{Department of Physics, University of Warwick, Coventry, CV4 7AL, UK.}
\author{K. G{\"o}tze}
\affiliation{Department of Physics, University of Warwick, Coventry, CV4 7AL, UK.}
\author{A. Szab{\'o}}
\affiliation{T.C.M. Group, Cavendish Laboratory, J. J. Thomson Avenue, University of Cambridge, Cambridge, CB3 0HE, UK.}
\author{T. S. Sikkenk}
\affiliation{T.C.M. Group, Cavendish Laboratory, J. J. Thomson Avenue, University of Cambridge, Cambridge, CB3 0HE, UK.}
\affiliation{Institute for Theoretical Physics and Center for Extreme Matter and Emergent Phenomena, Utrecht University, Leuvenlaan 4, 3584 CE Utrecht, The Netherlands.}
\author{M. R. Lees}
\affiliation{Department of Physics, University of Warwick, Coventry, CV4 7AL, UK.}
\author{A. T. Boothroyd}
\affiliation{Department of Physics, University of Oxford, Clarendon Laboratory, Oxford, OX1 3PU, UK.}
\author{D. Prabhakaran}
\affiliation{Department of Physics, University of Oxford, Clarendon Laboratory, Oxford, OX1 3PU, UK.}
\author{C. Castelnovo}
\email{cc726@cam.ac.uk}
\affiliation{T.C.M. Group, Cavendish Laboratory, J. J. Thomson Avenue, University of Cambridge, Cambridge, CB3 0HE, UK.}
\author{P. A. Goddard}
\email{p.goddard@warwick.ac.uk}
\affiliation{Department of Physics, University of Warwick, Coventry, CV4 7AL, UK.}

\date{\today}

\maketitle

%%%%%%%%%%%%%%%%%%%%%%%%%%%%%%%%%%%%%%%%%%%%%%%%%%%%%%%%%%%%%%%%%%%%%%%%%%%%%%%%%%

\section{Mechanisms linking monopole density and resistance}

The elementary excitations of the 2I2O ground state manifold of spin ices are effective sources or sinks of the (coarse-grained) magnetic field. According to the dumbbell model~\cite{Castelnovo08}, the magnetic field of a pyrochlore spin ice is a combination of the Coulombic field of these monopoles and comparatively short-ranged quadrupolar corrections. Therefore, the magnetic scattering of electron spins off this field pattern is approximated well by charge--dipole scattering off the monopoles only. 
Furthermore, lattice distortions due to the frustrated magnetic structure generate effective electric dipoles on each pyrochlore tetrahedron hosting a monopole~\cite{Khomskii12}, which in turn results in \textit{electric} charge--dipole scattering of the conduction electrons. 
It is reasonable to assume that these are the two dominant effects of the low-temperature holmium spin ice magnetism on the conduction electrons, and they are both charge--dipole type scattering off the emergent monopoles, in independent electric and magnetic channels.

Let us now estimate the scattering rate of conduction electrons due to these processes. The transition matrix elements for a generic charge--dipole interaction are given by the Fourier transform of the dipole potential:
\begin{equation}
    V(\v r) = -C\, \frac{\hat{\v p}\cdot \v r}{r^3} \implies V(\v q) = \int \ud\v r e^{-i\v q\cdot\v r} V(\v r) = 4\pi i C\, \frac{\hat{\v p}\cdot \v q}{q^2},
\end{equation}
where $\hat{\v p}$ is the unit vector parallel to the (electric or magnetic) dipole and $C$ is a generic coupling constant, equal to $ep/(4\pi\varepsilon_0)$ and $\mu_0\mu_\mathrm{B}q/(4\pi)$ in the electric and magnetic cases, respectively, where $q$ and $p$ are the magnetic charge and electric dipole moments of the monopoles. Now, the scattering rate follows from general scattering theory~\cite{Ashcroft76} as
\begin{align}
    \tau^{-1} &= \int \frac{\ud \v k'}{(2\pi)^3} W_{\v k,\v k'} (1-\hat{\v k}\cdot\hat{\v k}'),
    \label{eq: scattering rate isotropic}\\
    W_{\v k,\v k'} &= \frac{2\pi}{\hbar} n_\mathrm{mp} \:  \delta\left(\vphantom{\sum}E(\v k) - E(\v k')\right) \big| V(\v k'-\v k) \big|^2,
    \label{eq: scattering rate Boltzmann}
\end{align}
assuming the dispersion of conduction electrons is isotropic, and monopoles are dilute enough that electrons only scatter off one at a time. In Eq.~\eqref{eq: scattering rate Boltzmann}, $n_\mathrm{mp}$ is the number density of scatterers (monopoles), and the delta function enforces elastic scattering. 
To make progress, we make two further working assumptions: 
(i) the dispersion of conduction electrons is quadratic, $E(\v k) = \hbar^2k^2/(2 m)$, with an effective mass $m$ on the order of the bare electron mass; 
(ii) the orientation of dipoles, $\hat{\v p}$, is uniformly distributed on the unit sphere.
The latter is likely to hold for magnetic scattering, as the ordered iridium moments are small and their onset coincides with the metal--insulator transition, both of which suggest that the remaining conduction electrons are not spin-polarised. 
For electric scattering, the dipoles are oriented towards the minority spin of 3I1O and 1I3O tetrahedra~\cite{Khomskii12}; numerical evidence suggests that these are essentially uncorrelated, apart from the [111] field-polarised limit. 
Given these assumptions, Eqs.~(\ref{eq: scattering rate isotropic},\,\ref{eq: scattering rate Boltzmann}) can be evaluated to give
\begin{equation}
    \tau^{-1}(k) =  \frac{8\pi m^2C^2}{3\hbar^4 k^2}n_\mathrm{mp} v,
\end{equation}
where the group velocity is $v=\ud E/(\hbar\ud k) = \hbar k/m$. We note that the orientation of electric and magnetic dipoles in a given scattering process are likely uncorrelated (one belongs to the monopole, the other to the conduction electron), so the scattering rates due to these processes can simply be added (equivalently, the coupling constants $C$ are to be added in quadrature).
The correction to the resistivity can now be estimated from the Drude model:
\begin{equation}
    \Delta\rho = \frac{m}{n_\mathrm{e} e^2} \langle \tau^{-1}\rangle = \frac{4\pi m^2 C^2 n_\mathrm{mp}}{\hbar^3 k_\mathrm{F}n_\mathrm{e} e^2},
    \label{eq: resistivity final}
\end{equation}
where $n_\mathrm{e}$ is the number density of conduction electrons and $k_\mathrm{F} = (3\pi^2 n_\mathrm{e})^{1/3}$ is the Fermi wave vector assuming a quadratic dispersion. 

We can use Eq.~\eqref{eq: resistivity final} to estimate the carrier density in \hio.
The magnetic charge of an emergent monopole is $q = 2\mu_\mathrm{Ho}/a_\mathrm{d} \approx 4\times 10^{-13}\ \mathrm{Am}$~\cite{Castelnovo08},
while the moment of the corresponding dipole has been estimated to be on the order of $p\sim10^{-4} e\mathrm{\AA}$ in Dy$_{2}$Ti$_{2}$O$_{7}$~\cite{Lin15}.
It follows that the coupling constant $C$ is on the order of $1\ \mathrm{meV\AA^2}$ for both electric and magnetic scattering. 
The experimentally measured resistance hysteresis at $T=2$~K is about $\delta R\approx5~\mathrm{m\Omega}$ at its widest point; given the sample size $\ell\approx200~\mathrm{\mu m}$, this corresponds to a resistivity hysteresis $\delta\rho\approx 10^{-6}~\mathrm{\Omega\, m}$. The corresponding width of the monopole density hysteresis in our simulations is $\delta n_\mathrm{mp} \approx 5~\mathrm{nm^{-3}}$. 
These figures are consistent with a carrier density $n_\mathrm{e}\sim 10^{17}/\mathrm{cm}^3$ in~\eqref{eq: resistivity final}: this is some six orders of magnitude below the carrier density of elemental metals, a sensible figure for a badly conducting semimetal like \hio. 

It is important to point out that the metallic Fermi surface used in this estimate is an oversimplification of the electronic structure of \hio. The experimentally observed very slow increase of resistivity below the metal--insulator transition suggests that the system is not a band insulator (especially compared to Nd\textsubscript{2}Ir\textsubscript{2}O\textsubscript{7}~\cite{Matsuhira11,Tian16}), but rather a (semi)metal or a heavily doped semiconductor~\cite{Ishikawa12}, or perhaps not even a Fermi liquid~\cite{Wang20}. Nevertheless, we believe that our order of magnitude estimate remains reasonable even with these caveats in mind.

\clearpage

\section{Applied magnetic field parallel to [100]: full temperature dependence}

\begin{figure}[h]
\centering
\vspace{-5mm}
\includegraphics[width=0.52\textwidth]{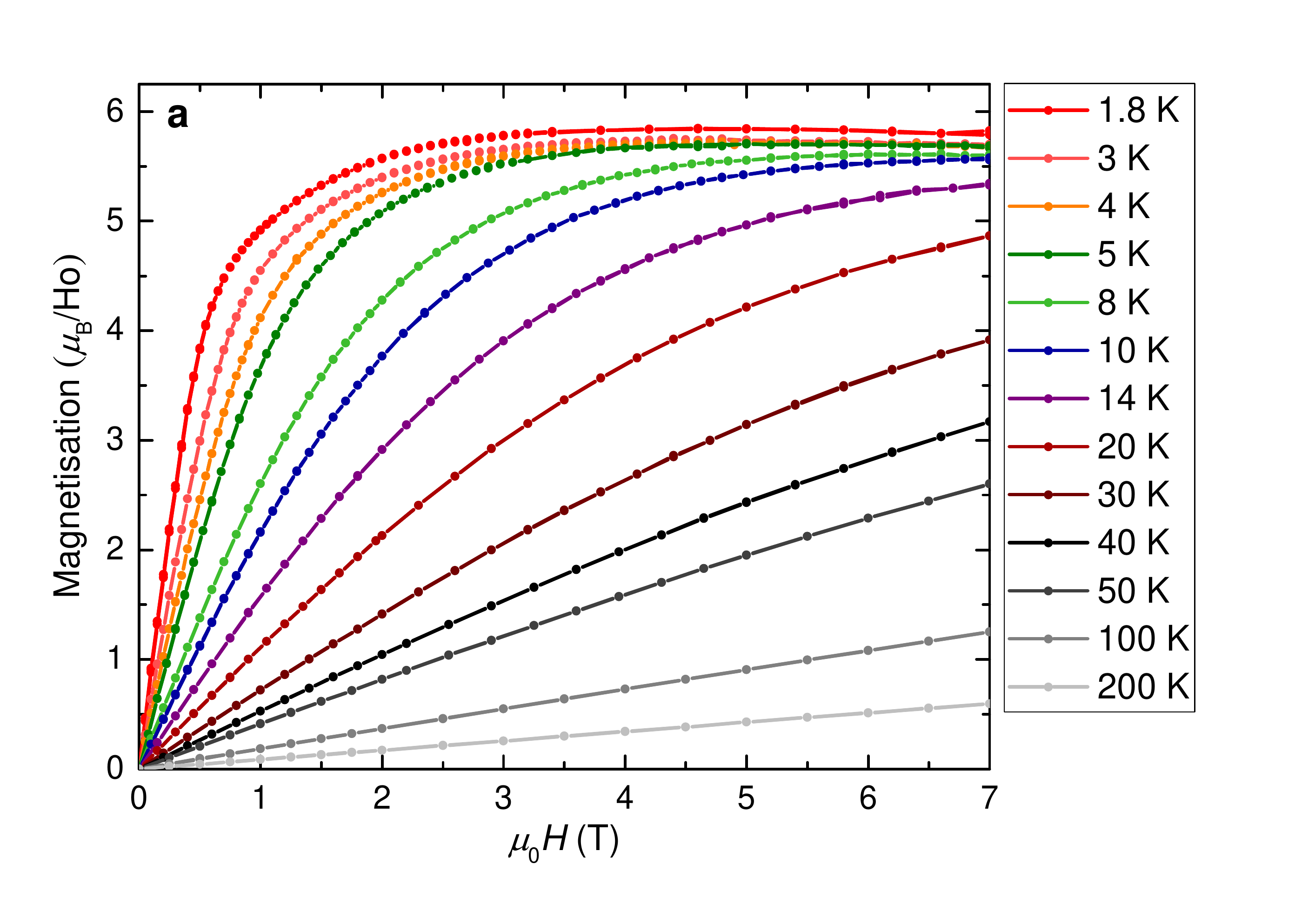}
\hspace{-10mm}
\includegraphics[width=0.52\textwidth]{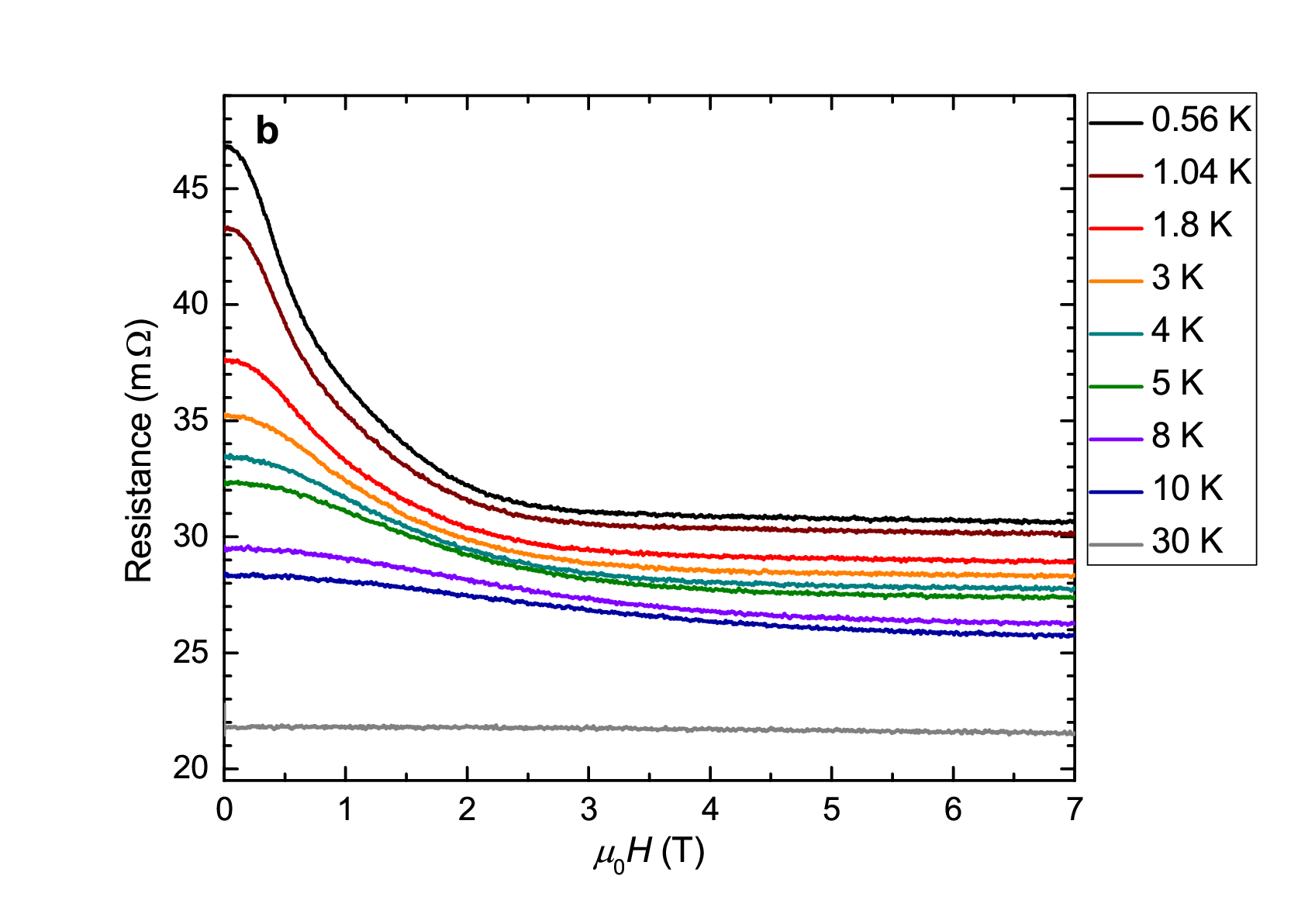}
\caption{Measurements of (a) the magnetisation and (b) the resistance of Ho$_{2}$Ir$_{2}$O$_{7}$ under an applied [100] magnetic field at various temperatures. In the main text the evolution of the magnetisation and resistance as the temperature is increased from 1.8~K to 10~K is discussed; this behaviour continues for the higher temperature data presented here.}
\label{M001SupplInf}
\vspace{0mm}
\end{figure}

\clearpage

\section{Applied magnetic field parallel to [111]: full temperature dependence}

\begin{figure}[h]
\centering
\includegraphics[width=\textwidth]{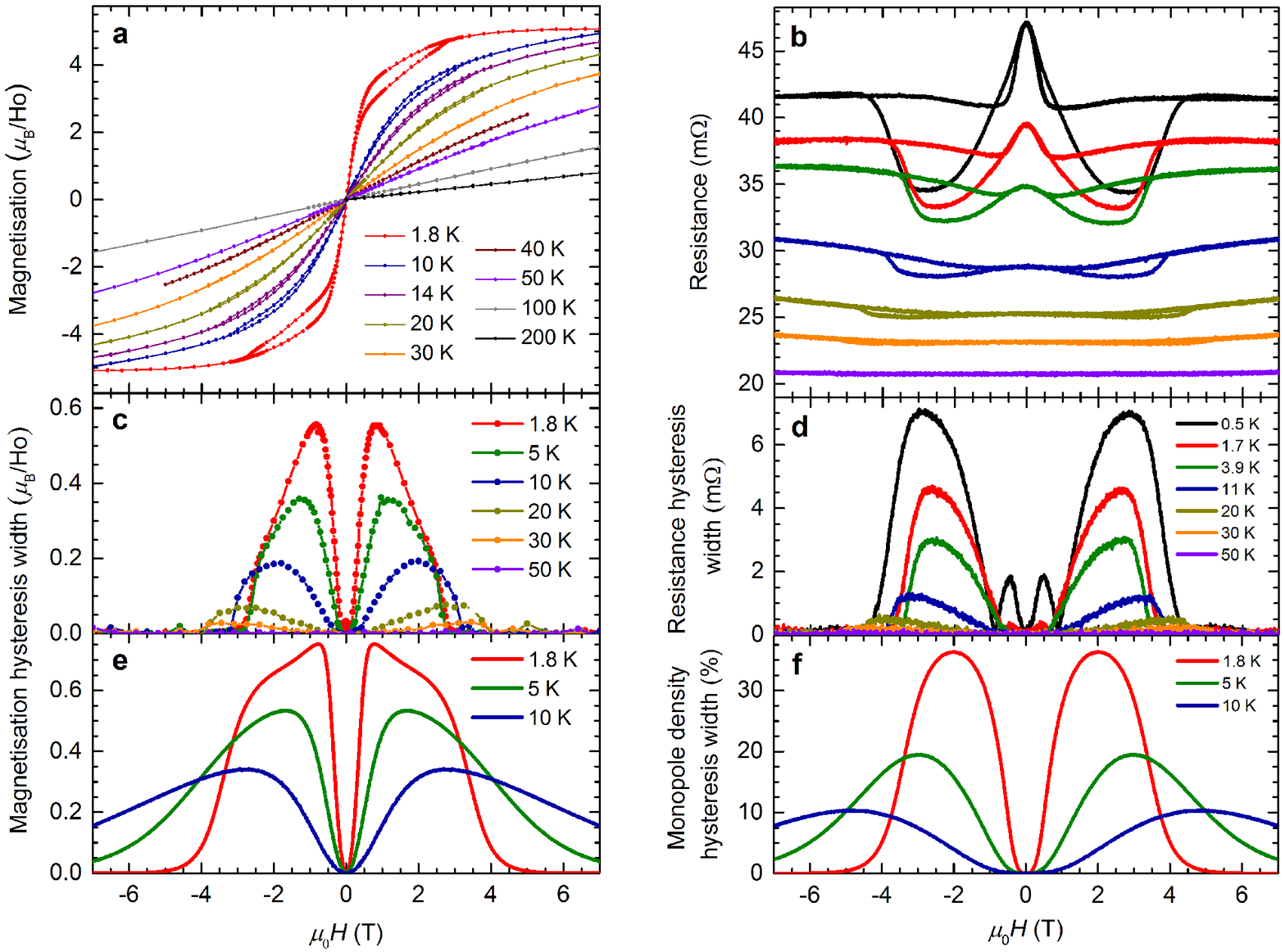}
\caption{Measurements of (a) the magnetisation and (b) the resistance of Ho$_{2}$Ir$_{2}$O$_{7}$ under an applied [111] magnetic field at various temperatures. The initial field sweep has been omitted from all traces for clarity. The width of the hysteresis, defined as the absolute value of the difference between the data measured on increasing and decreasing fields, is shown for a selection of temperatures for measurements of (c) the magnetisation and (d) the resistance and for Monte Carlo simulations of (e) the magnetisation and (f) the monopole density. The simulated hysteresis widths are calculated using a 30:70/70:30 ratio of Ir domains for -7~T to 7~T/7~T to -7~T, respectively, as this ratio gives the best agreement of the magnitude of the hysteresis width with experiments. The shape of the experimental hysteresis width is well reproduced by the simulations. We note the possibility that the resistance measurement at 0.5~K may be out of equilibrium; however we include it here for completeness.}
\label{TDep}
\vspace{0mm}
\end{figure}

\clearpage

\section{Variable field sweep rate magnetisation measurements}

\begin{figure}[h]
\centering
\vspace{-5mm}
\includegraphics[width=0.6\textwidth]{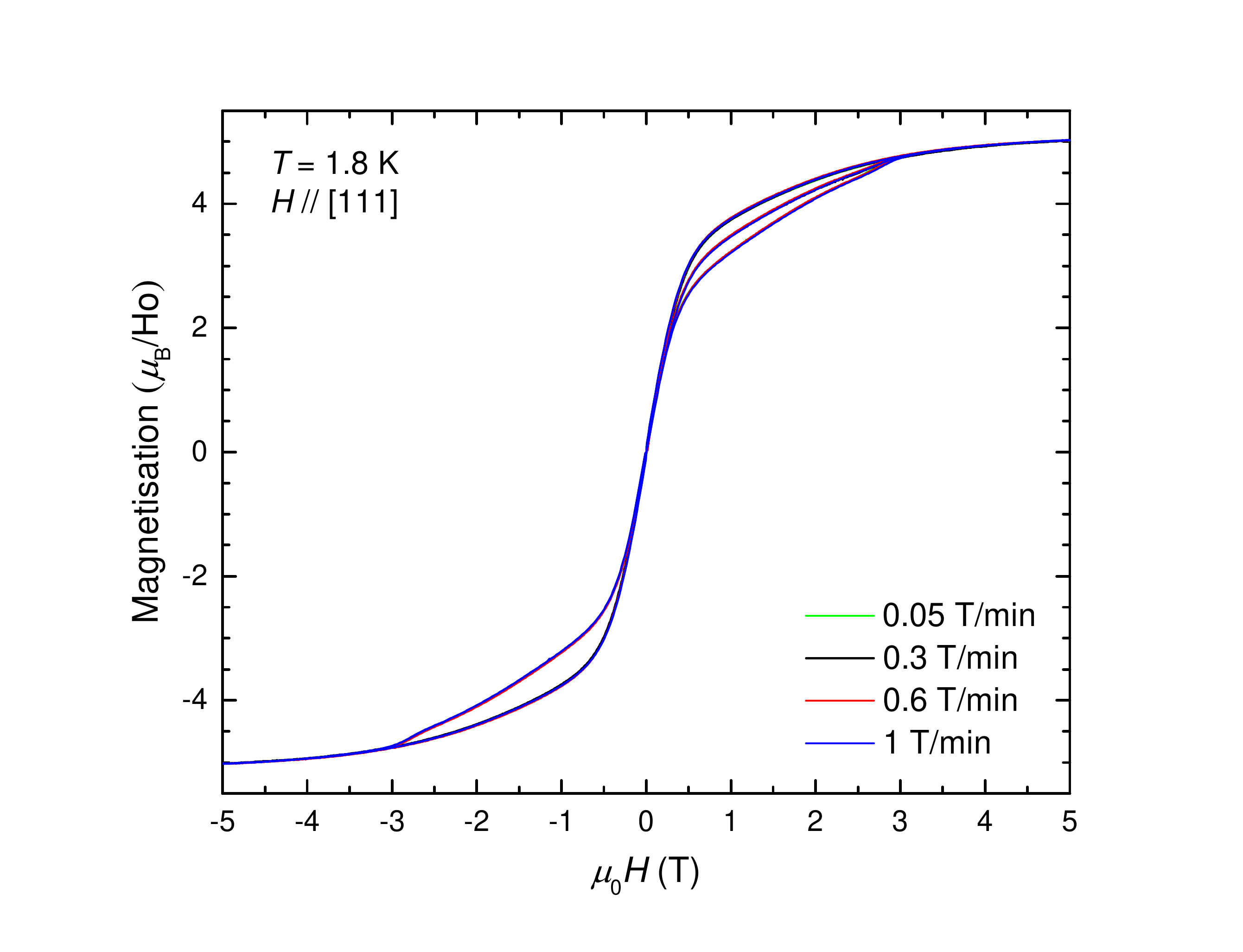}
\caption{Magnetisation of Ho$_{2}$Ir$_{2}$O$_{7}$ at 1.8~K under an applied [111] magnetic field swept at various rates. The hysteresis is insensitive to the sweep rate of the magnetic field across two orders of magnitude. Within experimental uncertainty, there is no significant broadening of the hysteresis as the sweep rate is increased, nor does it begin to open at zero applied field. This indicates that the hysteresis is static on the timescales of our experiments. A similar insensitivity to the magnetic field sweep rate was found for measurements of the magnetoresistance.}
\label{M111Rate}
\vspace{0mm}
\end{figure}

\clearpage

\section{Truncated magnetisation hysteresis loops}

\begin{figure}[h]
\vspace{-7mm}
\hspace{-11mm}
\includegraphics[width=0.54\textwidth]{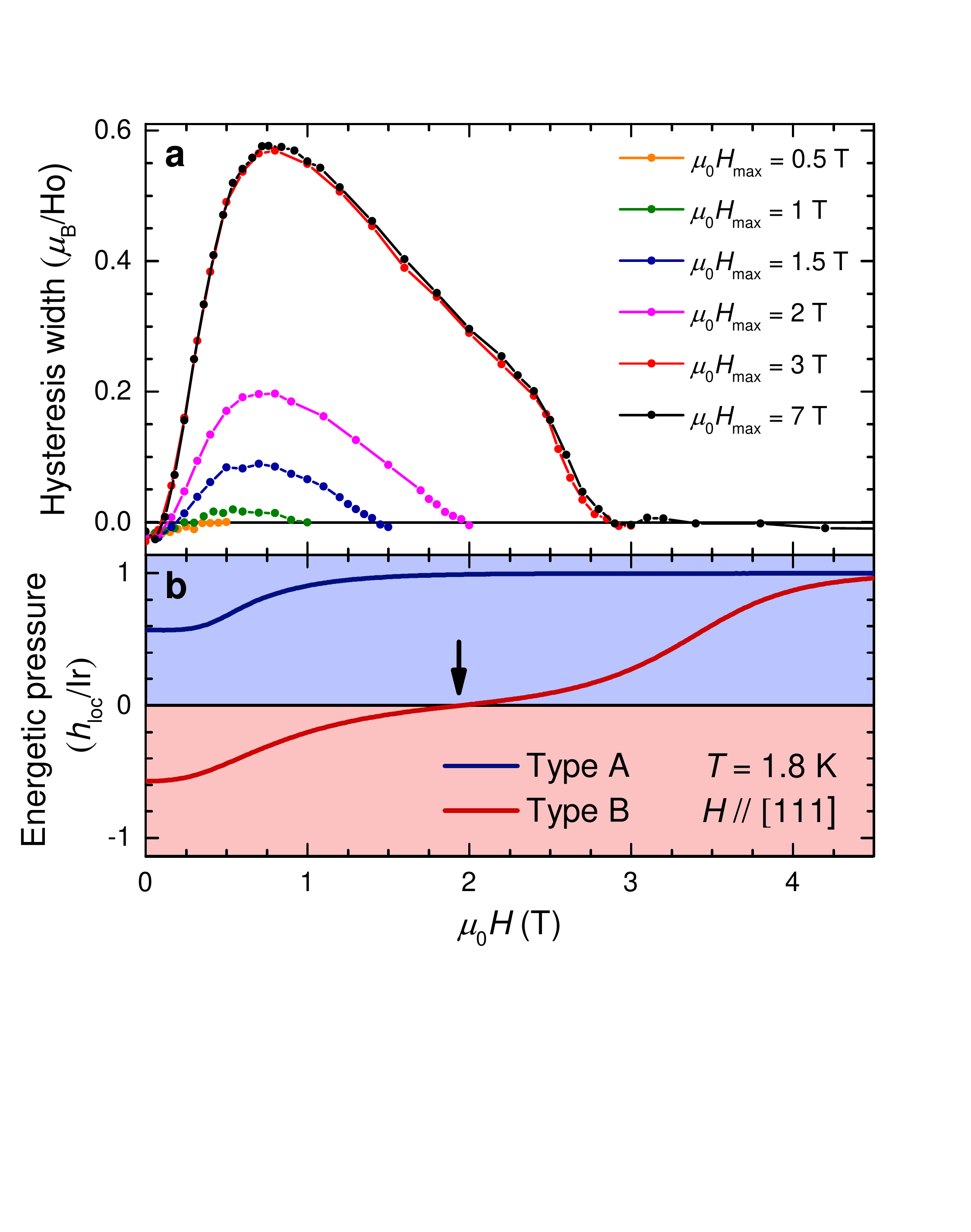}
\vspace{-20mm}
\caption{(a) Width of the hysteresis in the magnetisation $M(H)$ of Ho$_{2}$Ir$_{2}$O$_{7}$ under an applied [111] magnetic field at 1.8~K, defined here as the magnetisation measured on the downsweep minus data measured on the upsweep. For each magnetisation loop the magnetic field was swept from 0 to $H_{\rm max}$ to $-H_{\rm max}$ to $H_{\rm max}$ to 0 starting from a zero-field cooled initial condition, with each curve corresponding to a different $H_{\rm max}$. The virgin curve was not used in the calculation of the hysteresis width. The hysteresis width increases with $H_{\rm max}$ until $\mu_{\rm 0}H_{\rm max}$~$\approx$~3~T, beyond which it does not open any further. We note that the narrow region of negative hysteresis width as the applied magnetic field is swept through 0~T is an experimental artefact arising due to flux pinning in the superconducting magnet used to provide the magnetic field. (b) The simulated energetic pressure applied by the Ho moments in equilibrium inside both type~A and type~B Ir domains for $H\parallel$~[111] at 1.8~K. The background shading indicates which domain type the energetic pressure favours: A (light blue) or B (light red). For external fields below $\approx$ 2 T (arrow), both domains remain metastable and type A domains can grow at the expense of type B ones only by (slow) thermal fluctuations of the Ho moments over an energy barrier. This is consistent with the narrow experimental hysteresis loops observed over this range of field values in (a). Above 2~T, it becomes favourable for type B domains to flip, consistent with the larger hysteresis width above $\mu_{\rm 0}H_{\rm max}$~$\approx$~2~T in (a). We note also that the experimental hysteresis loops close well before the energetic pressure saturates, and in (a) once $\mu_{\rm 0}H_{\rm max}$ exceeds $\approx$~3~T the hysteresis width does not increase any further. This suggests that the domain-wall pinning overcome by the Ho-mediated energetic pressure is due to relatively weak pinning sites. We speculate that the presence of stronger pinning sites limits the maximum domain imbalance that can be reached to approximately 70:30. See Supplementary Section S7 for further discussion.} 
\label{Hyst}
\end{figure}

\clearpage

\section{Field-cooled magnetisation and resistance measurements}

\begin{figure}[h]
\centering
\vspace{-5mm}
\includegraphics[width=0.545\textwidth]{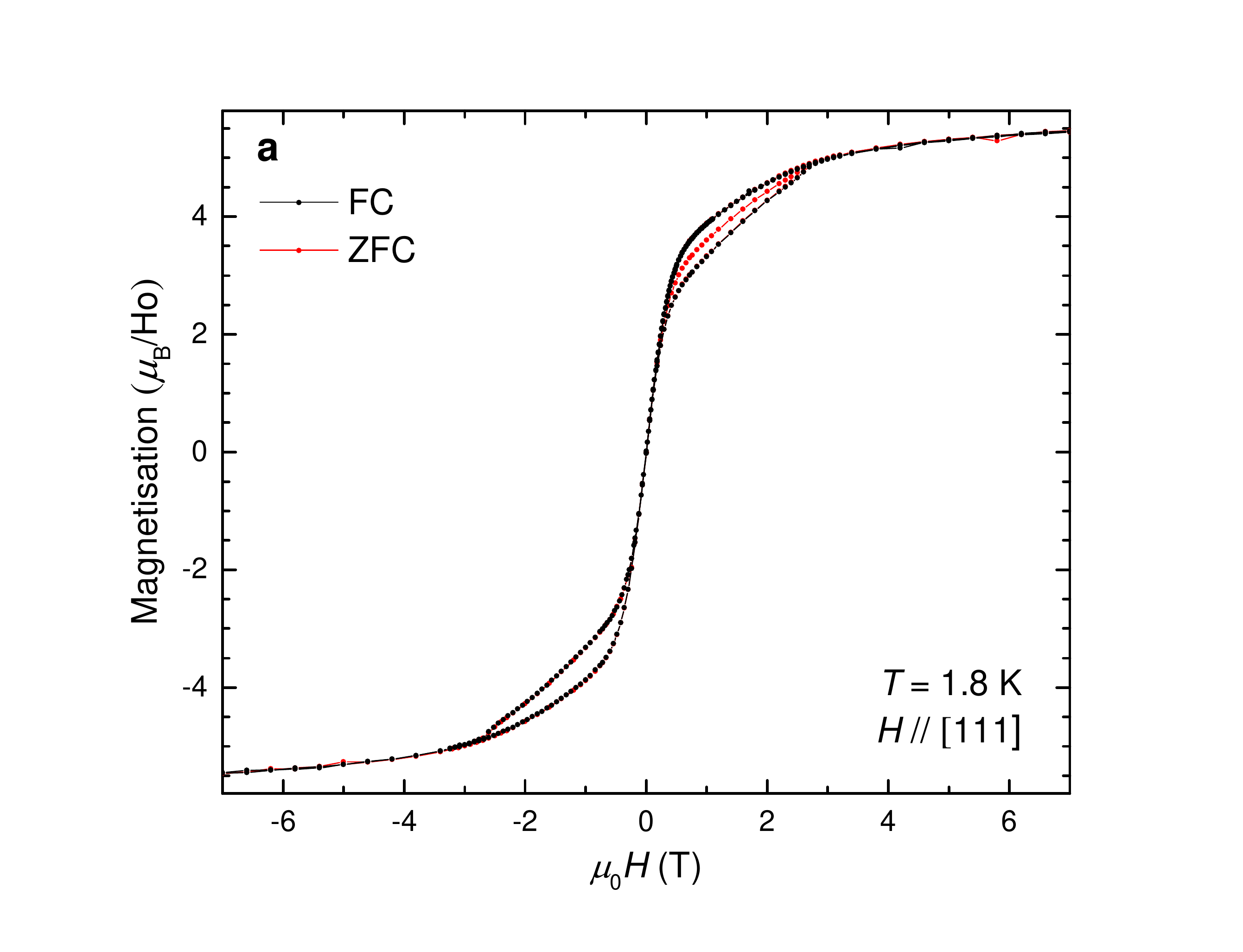}
\hspace{-19mm}
\includegraphics[width=0.545\textwidth]{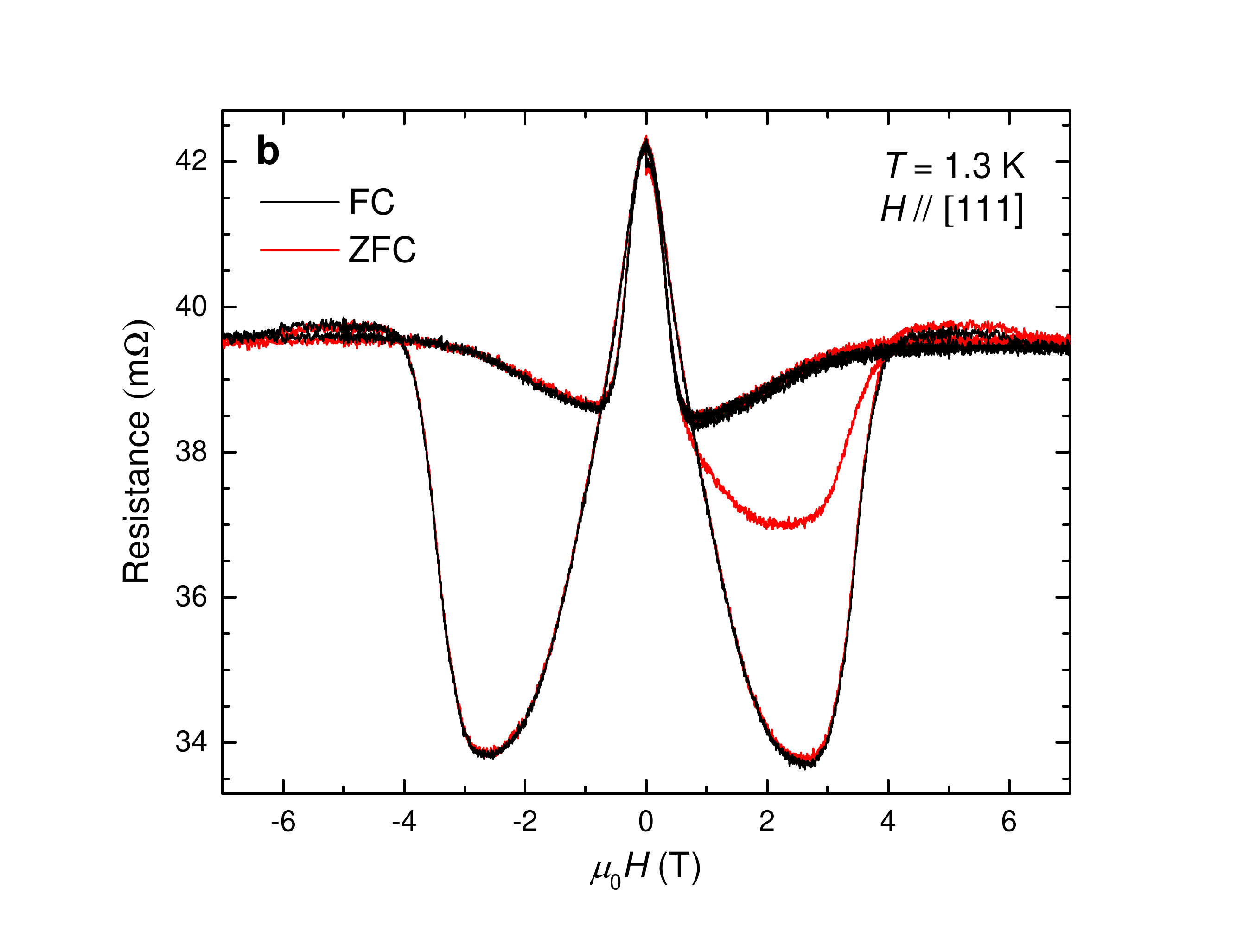}
\caption{(a) Magnetisation and (b) resistance of Ho$_{2}$Ir$_{2}$O$_{7}$ under an applied [111] magnetic field measured from a field-cooled (FC) and zero-field cooled (ZFC) initial condition. For the ZFC curves, the sample was cooled in zero applied magnetic field from 250~K/118~K prior to the commencement of the magnetisation/resistance measurements. For the FC curves, the sample was cooled in a 7~T/8~T [111] magnetic field from 250~K/54~K; once at 1.8~K/1.3~K the magnetic field was returned to 0~T and then the FC measurements were made. For both the magnetisation and resistance, the FC and ZFC measurements are identical within experimental uncertainty, except for the data recorded on the initial sweep to positive field. After cooling the sample in zero magnetic field, the two Ir domains form in an approximately equal ratio, and consequently data for the ZFC initial sweep follow the trajectory expected for a 50:50 domain ratio, passing through the middle of the subsequent downsweeps and upsweeps of the hysteresis loop (as is discussed in the main text, see Figure 3). By contrast, the FC initial sweep lies on top of the subsequent downsweep. This is consistent with the interpretation presented in the main text: upon cooling in a sufficiently large [111] magnetic field, the domain ratio will be altered plastically to a value estimated to be close to 70:30 by comparison to calculations. It will remain at this value once the field has been turned off, and consequently for the FC initial sweep the magnetisation and resistance follow the trajectory expected for a 70:30 domain ratio, rather than 50:50.}
\label{M111ZFCFC}
\vspace{0mm}
\end{figure}

\clearpage

%%%%%%%%%%%%%%%%%%%%%%%%%%%%%%%%%%%%%%%%%%%%%%%%%%%%%%%%%%%%%%%%%%%%%%%%%%%%%%%%%%

%
%
%%%%%%%%%%%%%%%%%%%%%%%%%%%%%%%%%%%%%%%%%%%%%%%%%%%%%%%%%%%%%%%

\section{Considerations about domain wall pinning and driving}

\subsection{Driving mechanism and pinning energy scales} 

In our work, we found evidence of plastic behaviour of the antiferromagnetic Ir domains on the time scales of our experiments. 
We propose that this domain wall movement is driven by their coupling to the Ho moments, specifically via the energy difference that a [111] saturated Ho configuration induces between the Ir type A and B domains. 
Within each domain, this energy difference is $2h_\mathrm{loc}\langle \sigma\rangle$ per Ir ion, 
where $\langle \sigma \rangle$ is the thermodynamic expectation value of the average of all Ho Ising variables $\sigma_i$ in equation~(1) of the main text, 
since the Ho--Ir interaction energy in type A and B domains is $\pm h_\mathrm{loc}\sum_i\sigma_i$, respectively.
This energy difference is plotted in Figure~S4.
For small external fields, $\langle\sigma\rangle$ is controlled by the Ho--Ir coupling and, as such, it favours the existing domain locally. 
While type A domains become energetically favourable in equilibrium for arbitrarily small positive (in our labelling convention) fields, 
they can only grow if holmium moments rearrange inside a type B domain to favour them locally:
for small fields this only happens by slow thermal fluctuations of the moments, which slows down domain wall movement substantially.
Above a certain field (around 2~T at 1.8~K), however, $\langle\sigma\rangle$ also becomes positive inside type B domains: after this point, the growth of type A domains is energetically favoured everywhere and so it can occur via ultrafast Ir dynamics, only hindered by domain wall pinning.

Concurrently with this interpretation, our experiments show that the hysteresis starts to open significantly only when the external field exceeds 2~T (Figure~S4). 
It is also seen that the hysteresis closes around 2.9~T, well below the saturation of the energetic pressure that flips type B domains:
this suggests that the domain wall pinning overcome by the Ho-mediated energetic pressure is due to weak pinning sites (or potentially self-pinning of the complex magnetic structure),
with net pinning energies on the order of 1~K per Ir ion.
This also helps explain the absence of long tails in the experimental hysteresis curves:
the critical field increases only slightly with temperature (Figure S6), causing the experimental hysteresis curve to close abruptly at slightly higher fields.

Such a weak pinning, however, does not explain the fact that the ratio of Ir domains appears to saturate around 70:30 rather than 100:0.
We speculate that this is due to rare but strong pinning sites (e.g., impurities): these give rise to relaxation time scales longer than the experimental ones, which prevent the domain distribution from becoming any more polarised.
Further work is necessary to fully understand the microscopic origin of the plastic behaviour in these materials beyond this simple energetic argument.

\begin{figure}[h]
\centering
\vspace{-5mm}
\includegraphics[width=0.52\textwidth]{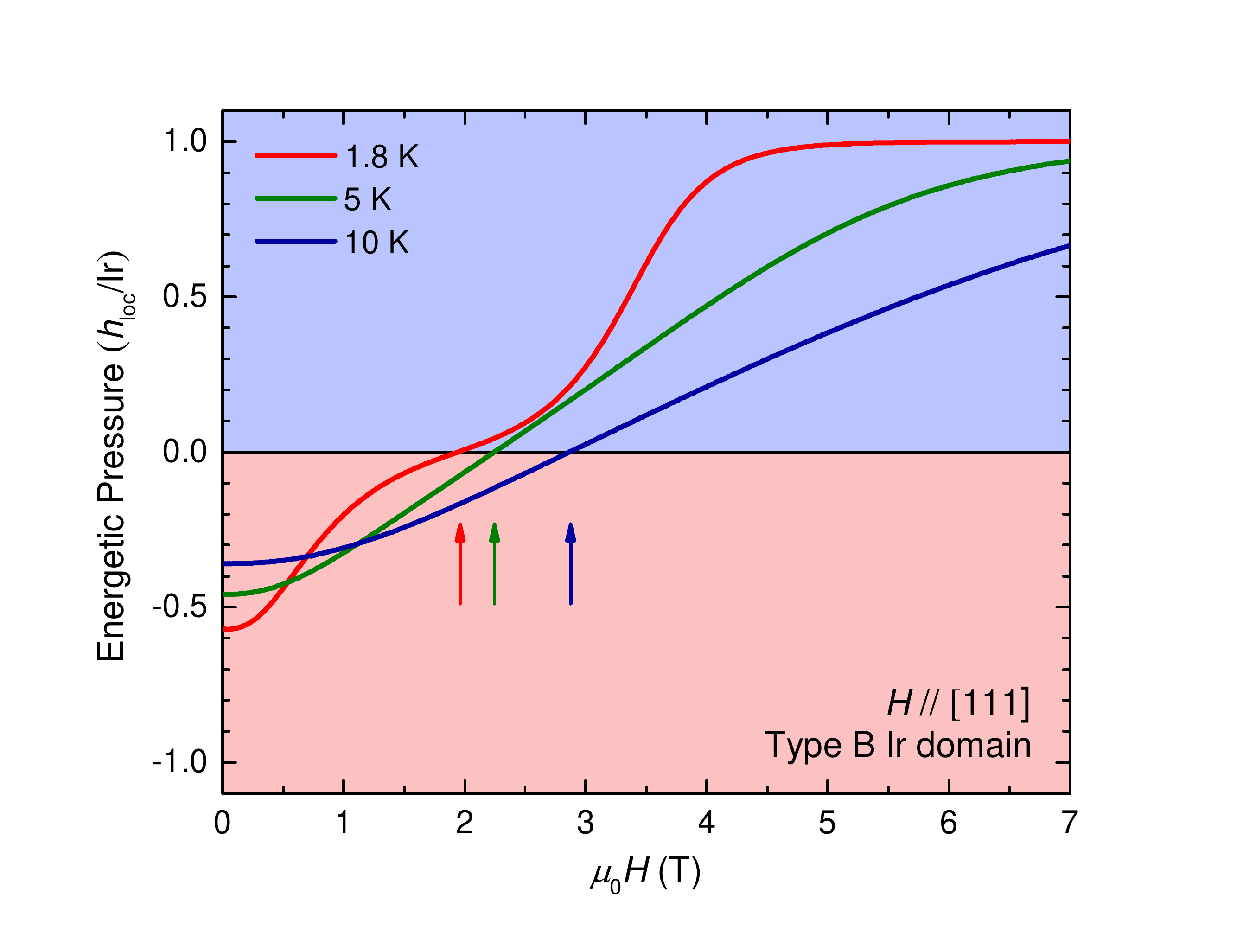}
\hspace{-10mm}
\caption{\textbf{Evolution of the energetic pressure with temperature}. Energetic pressure $2h_\mathrm{loc}\langle\sigma\rangle$ under an applied [111] magnetic field in type~B domains for three different temperatures. The background shading indicates which domain type the energetic pressure favours: A (light blue) or B (light red). The pressure changes sign, indicating a speedup in domain wall movement, at slightly increasing external fields (arrows). This trend matches the evolution of the field at which the experimental hysteresis loops close qualitatively (see Figure 3 and Figure S2).}
\label{}
\vspace{0mm}
\end{figure}

\subsection{Other sources of antiferromagnetic domain driving}

Earlier work claimed that some control over antiferromagnetic domains by an externally applied magnetic field in related systems could be achieved via the coupling of the field to the continuous canting of the spins~\cite{Arima13}, or via domain wall magnetisation induced by topological transport~\cite{Yamaji14} (see also Refs.~\cite{Tian16,Opherden17,Opherden18}). 
These mechanisms may be active in our system, leading to a direct coupling of the applied field to the iridium domain walls. However, their effect is substantially smaller than the one identified in our work (e.g., Ref.~\cite{Yamaji14} claims an effective uniform magnetisation of $10^{-3}\,\mu_\mathrm{B}$ per unit cell, which gives rise to an effective coupling strength on the order of millikelvins). 
As a result, even if these mechanisms were active in Ho$_{2}$Ir$_{2}$O$_{7}$, we are confident that the field--Ir coupling mediated by the Ho moments that we have identified in our work is dominant.

%%%%%%%%%%%%%%%%%%%%%%%%%%%%%%%%%%%%%%%%%%%%%%%%%%%%%%%%%%%%%%%%%%%%%%%%%%%%%%%%%%

\clearpage

\section{Applied magnetic field parallel to [110]}\label{Sec:110}

\begin{figure}[h]
\centering
\vspace{-5mm}
\includegraphics[width=0.52\textwidth]{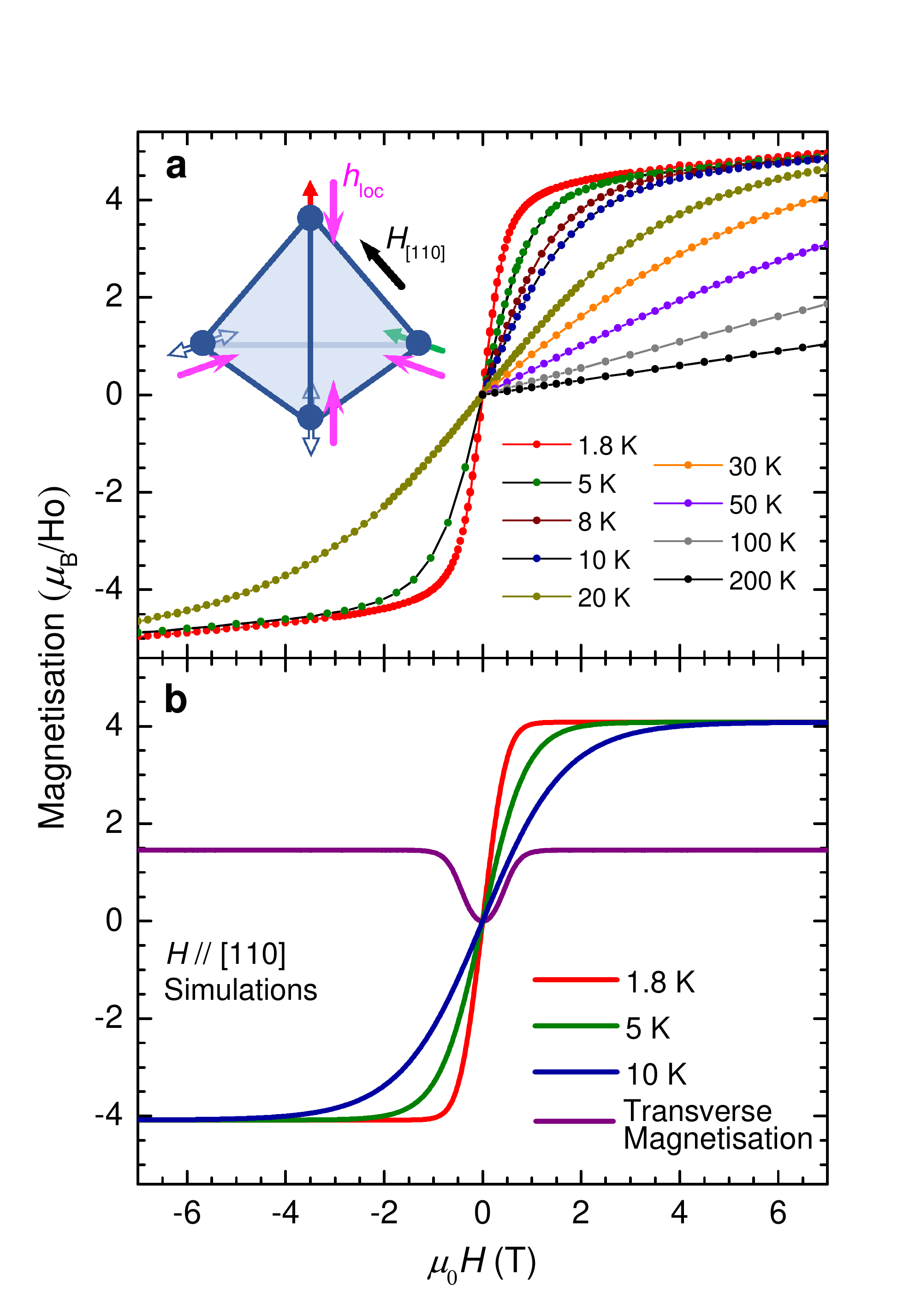}
\hspace{-10mm}
\includegraphics[width=0.52\textwidth]{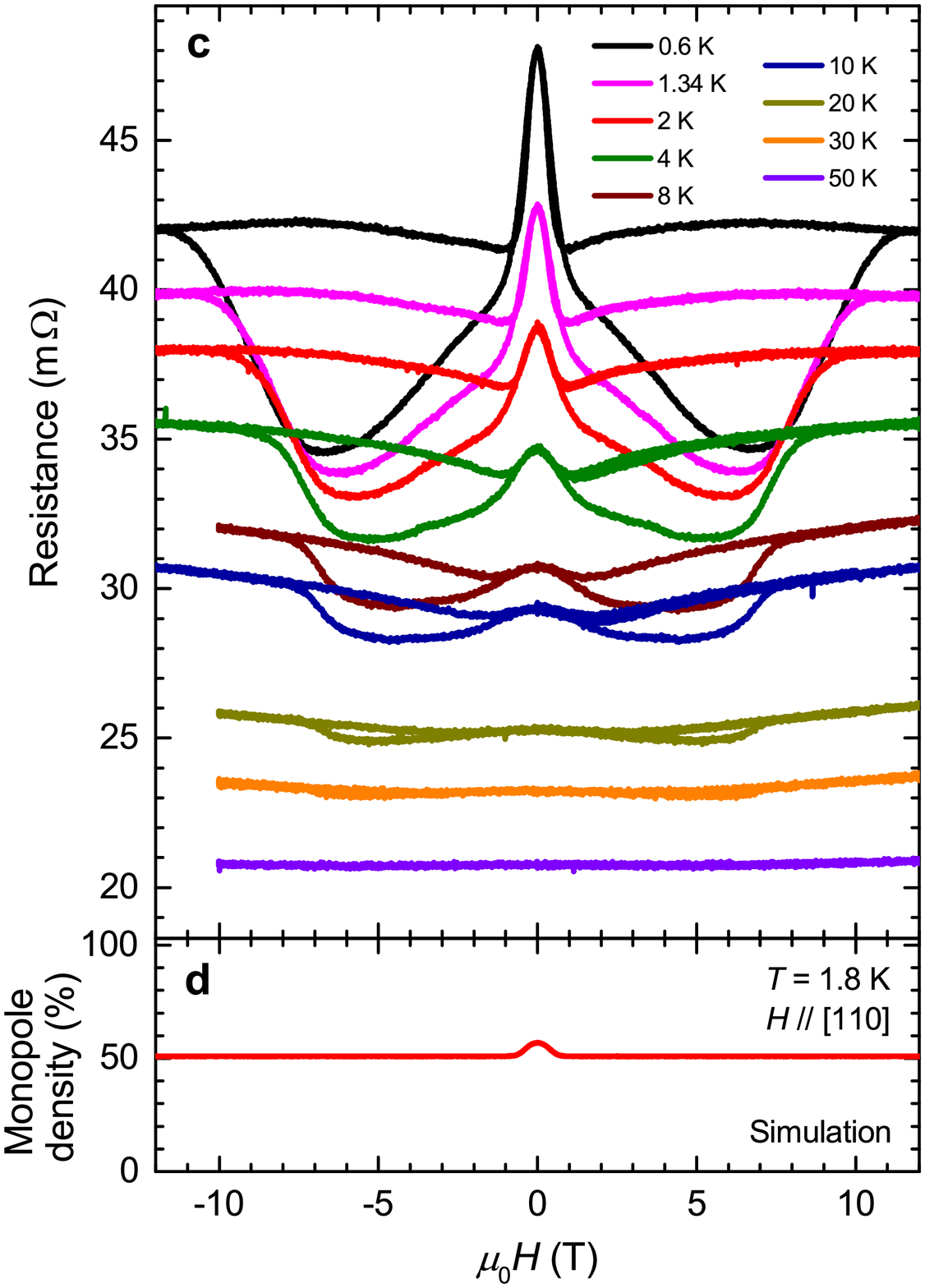}
\caption{\textbf{Ho$_{2}$Ir$_{2}$O$_{7}$ under the application of a [110] magnetic field}. (a) Measurements and (b) Monte Carlo simulations of the longitudinal magnetisation. The transverse magnetisation for a type~A domain (shown here for 1.8~K in the simulations only) is the component of the magnetisation along [001] whilst the field is applied along [110]. The inset to (a) shows a single tetrahedron of the Ho$^{3+}$ sublattice. Magenta arrows indicate the local effective field $\bf{h}_{\rm loc}$ due to the ordered Ir moments for a type-A domain. Under the application of an external [110] magnetic field (black arrow) one Ho moment orients parallel (green arrow) and one antiparallel (red arrow) to $\bf{h}_{\rm loc}$; the remaining two spins are normal to the [110] axis and thus are decoupled from the applied field. (c) Measurements of the resistance and (d) Monte Carlo simulations of the density of single monopoles. As with all other figures, demagnetisation effects have been accounted for as described in the methods section.}
\label{110Data}
\vspace{5mm}
\end{figure}

In this section, we present measurements of the magnetisation and magnetoresistance of Ho$_{2}$Ir$_{2}$O$_{7}$ under an applied [110] magnetic field which, whilst highly susceptible to misalignment, allow us to confirm some of the more subtle predictions and consequences of the theoretical interpretation presented in the main article.

The inset to Figure~\ref{110Data}a shows that upon applying a [110] magnetic field to a type-A Ir domain, one Ho moment in each tetrahedron orients parallel to $\bf{h}_{\rm loc}$, one antiparallel, and the remaining two spins do not couple to the external field as they are oriented perpendicular to [110]. An equivalent configuration is adopted in type-B Ir domains, but with the direction of $\bf{h}_{\rm loc}$ reversed. Consequently there is no net energetic pressure, and hysteresis is not expected for this orientation. 

We note that the two Ho moments which are decoupled from the external magnetic field can preferentially orient parallel to $\bf{h}_{\rm loc}$. While this has no effect on the longitudinal magnetisation, since the two spins are normal to [110], it does result in a net magnetisation along [001]. A Monte Carlo simulation of this transverse magnetisation ($M\parallel[001]$ for $H\parallel[110]$ at $T = 1.8$~K) for a type-A single-domain crystal is shown in Figure~\ref{110Data}b (the transverse magnetisation for a type-B domain is the same curve multiplied by a factor of $-1$). For a multidomain crystal the bulk transverse magnetisation is expected to be non-zero except in the case of a 50:50 domain ratio (for which the contributions from the two domain types cancel), and saturates at a magnetisation which depends on $\bf{h}_{\rm loc}$, the temperature, and the Ir domain ratio imbalance. Notably, this transverse magnetisation (due to the Ho moments) only occurs because of the magnetism of the ordered Ir moments (via $\bf{h}_{\rm loc}$), and so is not expected to be present in the analogue titanate compound Ho$_{2}$Ti$_{2}$O$_{7}$.

The expected absence of hysteresis for $H\parallel[110]$ is apparent in Monte Carlo simulations of the magnetisation (Figure~\ref{110Data}b) and the monopole density (Figure~\ref{110Data}d). The simulated magnetisation saturates at 4.08~$\mu_{\rm B}$/Ho, as is expected for the spin orientation described above and the monopole density is broadly insensitive to the applied magnetic field. However, experimental measurements of the resistance (Figure~\ref{110Data}c) show highly hysteretic behaviour which is similar in form to data measured under an applied [111] field (see main article), but closes at higher fields. The magnetisation (Figure~\ref{110Data}a), whilst non-hysteretic, exceeds the saturation magnetisation value expected for this orientation and indeed does not saturate over the measured field range, instead retaining a positive gradient at high fields.

\begin{figure}[t]
\centering
\vspace{-5mm}
\includegraphics[width=0.52\textwidth]{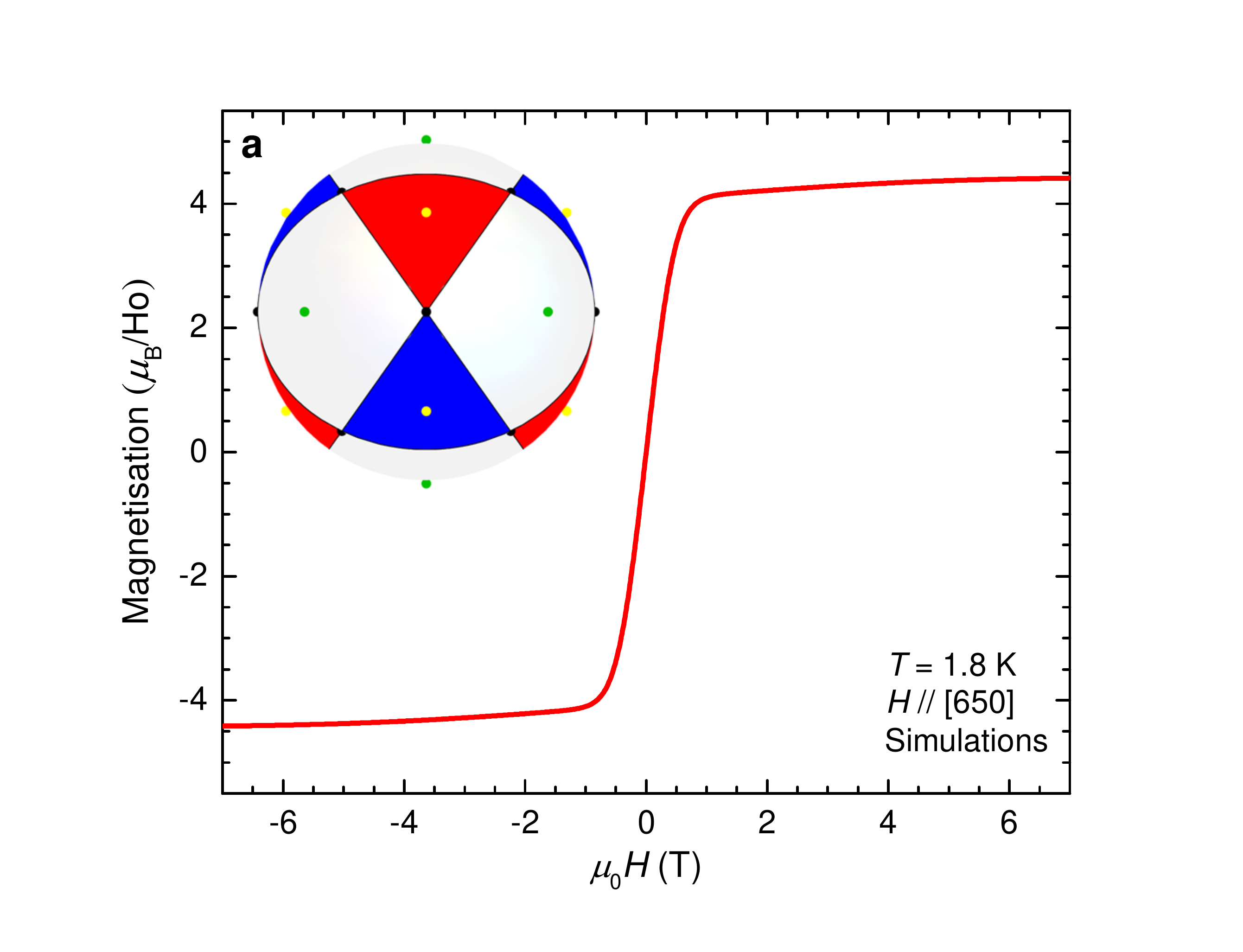}
\hspace{-10mm}
\includegraphics[width=0.52\textwidth]{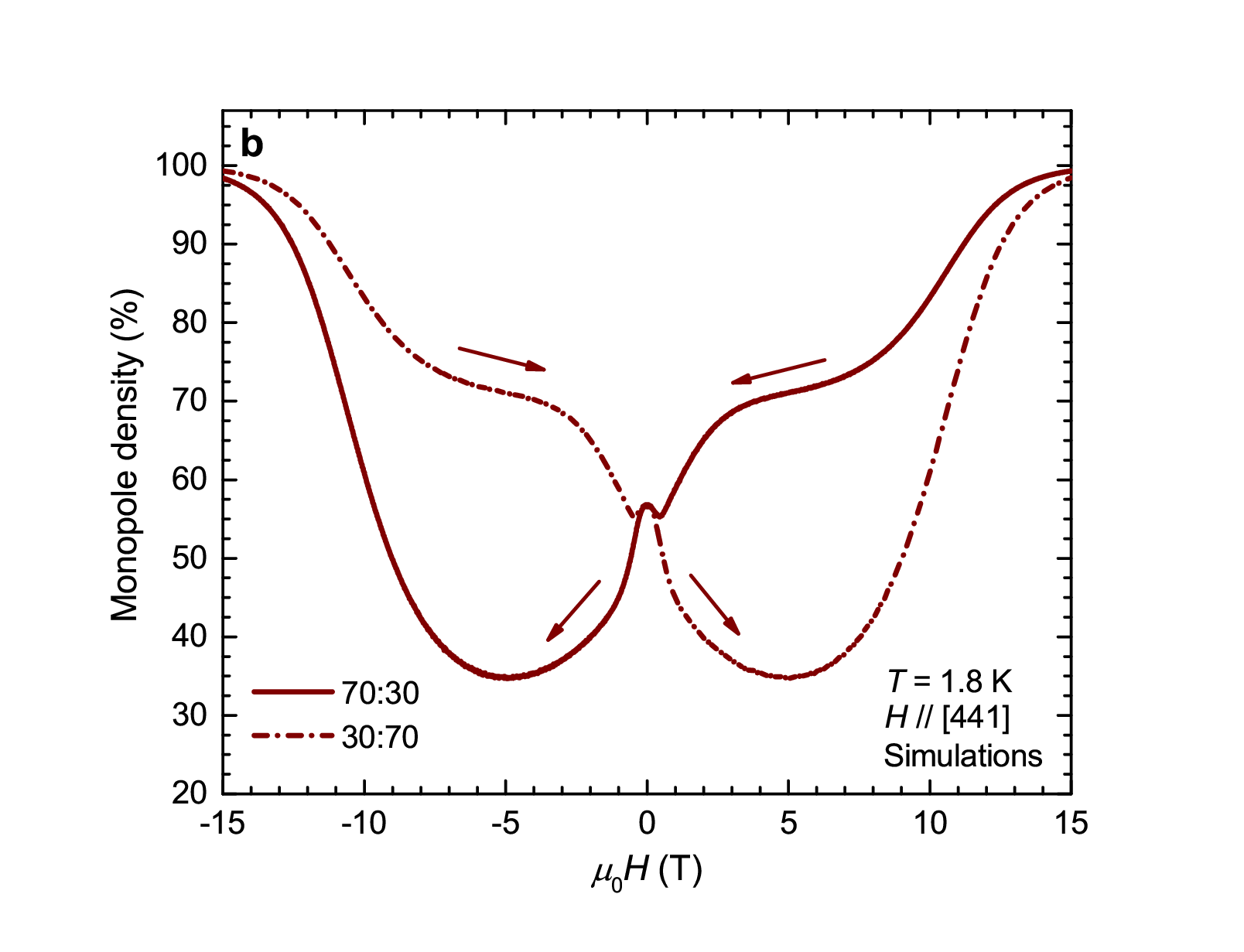}
\caption{\textbf{Tilted magnetisation and monopole density simulations.} Monte Carlo simulations at 1.8~K of (a) the magnetisation under an applied magnetic field along the [650] crystallographic direction, which corresponds to a tilt from [110] towards [100] of 5.19\degree and (b) the density of single monopoles under an applied field along [441], which corresponds to a tilt from [110] towards [111] of 10.02\degree. The monopole density is calculated using an A:B ratio of Ir-domain types of 70:30 for 15~T to -15~T and 30:70 for -15~T to 15~T. The resultant evolution of the hysteresis upon sweeping the field is indicated by the arrows. The tilts for the simulations were chosen as they provided a good qualitative agreement with the experimental data in Figures S7a and S7c; consequently they approximately correspond to the estimated misalignment of these measurements. The inset to (a) depicts the high-field polarised state of the Ho moments for different directions of the externally applied magnetic field, represented as a 2D projection of the applied field direction. The green, yellow, and black dots correspond to the $\langle$100$\rangle$, $\langle$111$\rangle$, and $\langle$110$\rangle$ directions, respectively. The colour coding of the regions identifies the corresponding lowest Zeeman energy state: the light grey regions are 2I2O configurations and the red and blue regions are 3I1O/1I3O monopole crystals.}
\label{110Tilt}
\vspace{0mm}
\end{figure}

The divergence of the experimental results from the expected behaviour for this field direction arises due to the particular sensitivity to misalignment of the measured properties close to the [110] orientation. The inset to Figure~\ref{110Tilt}a shows that the [110] order (where one spin points into the tetrahedron, one points out, and two are decoupled from the external field) is stabilised only when the applied field is perfectly aligned in that direction. Any deviation, no matter how small, couples the applied field to the two Ho spins per tetrahedron which are normal to the [110] direction. These spins can then polarise to produce a 2I2O configuration or a 3I1O/1I3O monopole crystal, depending on the precise direction of the field misalignment. This is in contrast to the situation for applied [100] and [111] fields, for which the respective 2I2O order and 3I1O/1I3O monopole crystals are stabilised over a wide range of field angles around the [100] and [111] directions.

Figure~\ref{110Tilt}a shows a Monte Carlo simulation of the magnetisation for an applied field tilted by 5.19\degree~away from [110] towards [100], which reproduces the experimentally observed behaviour well. There is an initial rapid rise in the magnetisation as the two spins which are not perpendicular to [110] are polarised by the dominant field component. At higher fields, the two spins unconstrained by the [110] component are polarised by the smaller [100] component to yield a 2I2O configuration, leading to a subsequent slower increase of the magnetisation and a saturation magnetisation which exceeds that expected for a perfectly aligned [110] field. Figure~\ref{110Tilt}b shows a Monte Carlo simulation of the monopole density for an applied field tilted by 10.02\degree~away from [110] towards [111]. The [111] component of the applied field couples to the two spins that lie perpendicular to [110] and the system orders into a 3I1O/1I3O monopole crystal, which generates an energetic pressure on the Ir domain walls. This plastically deforms the Ir domain ratio, opening a hysteresis similar to that observed for the [111] orientation, but closing at higher fields, because a smaller component of the applied field is directed along [111]. This behaviour of the monopole density reproduces the observed hysteresis in measurements of the resistance well.

\bibliography{SIPrePrint}